\documentclass{article}

\usepackage{arxiv}

\usepackage[utf8]{inputenc} 
\usepackage[T1]{fontenc}    
\usepackage{hyperref}       
\usepackage{url}            
\usepackage{booktabs}       
\usepackage{amsfonts}       
\usepackage{nicefrac}       
\usepackage{microtype}      
\usepackage{lipsum}
\usepackage{graphicx}
\graphicspath{ {./images/} }

\usepackage{algorithm}
\usepackage{algpseudocode}
\usepackage{mathrsfs}
\usepackage{lineno}
\usepackage{amsmath}
\usepackage{mathtools}

\title{A Fuzzy Cascaded Proportional-Derivative Controller for Under-actuated Flexible Joint Manipulators Using Bayesian Optimization}

\author{
 Changyi Lei \\
  King's College London\\
  London, UK \\
  \texttt{changyi.lei@kcl.ac.uk} \\
  \And
 Quanmin Zhu \\
  University of the West of England\\
  Bristol, UK \\
  \texttt{quan.zhu@uwe.ac.uk} \\
}

\begin{document}
\maketitle
\begin{abstract}
  This paper proposes a novel fuzzy cascaded Proportional-Derivative (PD) controller for under-actuated single-link flexible joint manipulators. The original flexible joint system is considered as two coupled $2^{nd}$-order sub-systems. The proposed controller is composed of two cascaded PD controllers and two fuzzy logic regulators (FLRs). The first (virtual) PD controller is used to generate desired control input that stabilizes the first $2^{nd}$-order sub-system. Solving the equation by considering the coupling terms as design variables, the reference signal is generated for the second sub-system. Then through simple compensation design, together with the second PD controller, the cascaded PD controller is derived. In order to further improve the performance, two FLRs are implemented that adaptively tune the parameters of PD controllers. Under natural assumptions, the cascaded fuzzy PD controller is proved to possess locally asymptotic stability. All the offline tuning processes are completed data-efficiently by Bayesian Optimization. The results in simulation illustrate the stability and validity of our proposed method. Besides, the idea of cascaded PD controller presented here may be extended as a novel control method for other under-actuated systems, and the stability analysis renders a new perspective towards the stability proof of all other fuzzy-enhanced PID controllers.
\end{abstract}


\section{Introduction}

Flexible-joint manipulators (FJM) represent a class of manipulators whose joint is made of flexible material. Compared with rigid-body manipulators, FJM requires small actuation, low energy consumption and low rate of damages \cite{Khoygani2015IntelligentNO,Fateh2012NonlinearCO}. Nonetheless, the introduction of flexibility also increases the complexity of control. In practice, FJM is a highly nonlinear, strongly coupled and time-varying system \cite{Rachidi2014ProportionalIntegralSM}. Besides, the degree-of-freedom (DoF) of FJM is larger than its number of control input. Consequently, research into the control method of FJM is valuable to the industry.

Due to its wider importance in industrial applications, flexible joint manipulator has been a heated point of research. Over decades, many control methods have been proposed. Yan et al. proposed a robust controller based on equivalent-input-disturbance (EID) \cite{Yan2020TrackingCO,Yan2021ANR}. An EID estimator, full-order state observer and state-feedback control law were realized to ensure the stability of the system.  In \cite{Dian2019AdaptiveBC}, an adaptive backstepping control method was presented, using Interval Type-2 Fuzzy Neural Network to estimate the unknown dynamics of the system. The errors of the system was proved to be bounded under Lyapunov sense. Yang et al. designed a cascaded controlling composed of three modules, namely an adaptive controller, a torque-tracking controller, and a motor controller \cite{Yang2020ANC}. A Kalman observer was used to estimate the state variables, torque as well as their higher-order derivatives. All controllers are designed based on Lyapunov stability theorem. Although the above mentioned methods have satisfying performance, they require too much human craftsmanship and increased computational cost. 

Among all the other existing controllers, Proportional-Integral-Derivative (PID) controller has been one of the most widely used in the industry. PID controller, though simple in calculation, has proved to be effective for a wide range of nonlinear systems in practice. Therefore, for flexible joint manipulators, many PID-based controllers have been proposed, trying to solve the problem in a simple but effective way. A neural network based PID controller was proposed to improve the performance of conventional PID \cite{Jie2021FlexibleJM}. A 3-layer feedforward neural network is used to output the parameters of PID controller online, and is trained using steepest descent algorithm to decrease tracking error. Results show that neural network based PID controller has lower tracking error and faster convergence speed. In \cite{Dehghani2015FuzzyLS} and \cite{Dehghani_khodadadi_2016}, PID controller is enhanced by a fuzzy logic system (FLS). The FLS takes the input of error vector and output adjustments to the parameters of PID controller. In terms of performance, the FLS enhaced PID has smooth tracking performance without overshoot.  A multi-PID controller scheme was proposed in \cite{Yang2021AMP}. The structure was composed of joint toque generator, torque tracker and motor position controller, which are all realized by simple PID. Besides, a friction observer is mounted to increase disturbance rejection. Note that most PID-based controllers proposed try to mount on other complicated modules, in order to increase the order of the controller. However, integrating fuzzy logic and neural network usually makes the systems performance intractable. Besides, these kinds of methods usually lack a systematic design approach and the stability analysis is hard to be derived, especially with nonlinearity and disturbances.

Therefore, in this paper, we propose a novel cascaded PD controller to solve the problem. The original $4^{th}$-order system is considered as two coupled $2^{nd}$-order dynamics (namely sub-plant1 and sub-plant2). Two PD controllers are used. The first PD controller is integrated with the coupling terms in sub-plant1, which calculates a reference signal for sub-plant2. In this way, the internal dynamics is tractable and can be specified. Then the second PD controller can just maintain the stability of sub-plant2. The total order of controller in this case is 4, which is equivalent to the order of the system. We believe the proposed method not only preserves the simplicity of PID controller, but also achieves certain degree of internal dynamics control. What is more, the stability analysis of the original nonlinear system with disturbance can be derived without any linearisation or simplification, which is challenging in previous PID controller research. Our method can achieve non-oscillating performance without mounting on any other modules. However, fuzzy logic system is still implemented to further improve the performance of cascaded PD controller, along with stability analysis using the lower bounds of the FLRs. To complete the task of controller tuning in a data efficient way, Bayesian Optimization (BO) is chosen.

The contributions of this paper are summarized as follows:
\begin{itemize}
    \item A novel cascaded PD controller for uncertain under-actuated $4^{th}$-order flexible joint systems is proposed.
    \item Type-1 fuzzy logic regulators are integrated into cascaded PD controllers, which shorten the settling time and further cancels oscillation.
    \item The asymptotic stability of proposed cascaded fuzzy PD controllers with respect to $4^{th}$-order system dynamics with uncertainties is proved under natural assumptions. This renders a new perspective for the stability analysis of other fuzzy-enhanced PID controllers.
    \item Bayesian Optimization is implemented for joint tuning of cascaded PD controllers, as well as fuzzy logic system tuning.
\end{itemize}

The rest of the paper is organized as follows. Firstly, some preliminaries are introduced, including dynamic model to be investigated, fuzzy PD controller, and Bayesian Optimization rationale. Secondly, the controller is meticulously designed and analyzed. The cascaded PD controller is designed step by step, and then the fuzzy logic regulator is integrated. Afterwards, the stability analysis is given in terms of transfer function and Jacobian matrix. Sec.\ref{simulation} presents the simulation results. The parameters are specified and BO tuning is executed to determine all the variables. Then the asymptotic stability is proved, with numerical simulation results of tracking tasks. In addition, several runs of ablation experiments are conducted, which reveals the advantages and potential challenges of proposed method. Sec.\ref{conclusion} concludes the paper and points out future research directions.

\section{Preliminaries}

\subsection{Dynamic Model Description}
\begin{figure}[thpb]
  \centering
  \includegraphics[scale=0.5]{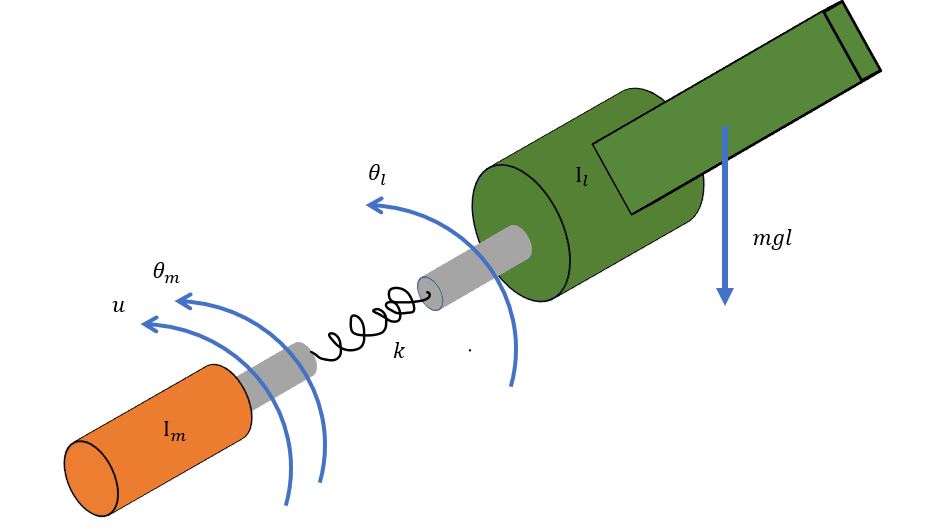}
  \caption{Conceptual structure of flexible joint manipulator}
  \label{manipulator}
\end{figure}

Fig.\ref{manipulator} shows the conceptual structure of a single-link flexible joint manipulator (SLFJM). It is composed of two solid bodies, namely the motor shaft and the link, whose connection is modeled as a torsion spring. In the figure, $I_m$ and $I_l$ are the inertia of the motor and the link respectively. $\theta_m$ is the rotational angle of the motor, and $\theta_l$ is correspondingly that of the link. $u$ represents our control torque input. $k$ is the torsional coefficient of the spring, $m$ is the mass of the link, $g$ is the gravity coefficient and $l$ is the shortest distance between the center of mass (CoM) of the link to the rotational axial. The rationale of SLFJM is like this: the input $u$ will cause a discrepancy between $\theta_m$ and $\theta_l$, which will generate torsion due to the torsional spring. The induced torque is proportional to $|\theta_m-\theta_l|$. 

Based on the above discussion and Newton's Law, a simplified nominal dynamic model of SLFJM can be established as \cite{Yan2021ANR}
\begin{equation}
\begin{cases}
I_l \ddot{\theta}_l+mglcos\theta_l +k(\theta_l-\theta_m)=0 \\
I_m \ddot{\theta}_m+\mu \dot{\theta}_m -k(\theta_l-\theta_m)-u=0,
\end{cases} \label{dynamicmodel1}
\end{equation}
where $\mu$ is the friction coefficient, $\dot{\theta}_l, \dot{\theta}_m$ are angular velocities of the link and motor, $\ddot{\theta}_l, \ddot{\theta}_m$ are angular accelerations of the link and motor, $u$ is the control input. For convenience and generality of expression, we replace the above variables with 
\begin{equation}
\begin{cases}
x_1 = \theta_l \\
x_2 = \dot{\theta}_l \\
x_3 = \theta_m \\
x_4 = \dot{\theta}_m \\
\end{cases} \label{dynamicmodel2}
\end{equation}
Rearranging (\ref{dynamicmodel1}) and (\ref{dynamicmodel2}) into state-space equation, and introducing total disturbance $d_1,d_2$, we have 
\begin{equation}
\begin{cases}
\dot{x}_1 = x_2 \\
\dot{x}_2 = -\frac{mgl}{I_l}cosx_1-\frac{k}{I_l}(x_1-x_3)+d_1 \\
\dot{x}_3 = x_4 \\
\dot{x}_4 = \frac{k}{I_m}(x_1-x_3)-\frac{\mu}{I_m}x_4 + \frac{1}{I_m}u+d_2. \\
\end{cases} \label{dynamicmodel3}
\end{equation}

Note that system (\ref{dynamicmodel3}) is an under-actuated system with 2 degree-of-freedom (DoF) but only 1 control input. Besides, it is noticeable that this system has fourth-order dynamics, and can be considered as two second-order systems being connected by a torsional spring. These two facets determine the difficulty of controlling the system.

\subsection{Fuzzy PD Controller}

\subsubsection{Conventional PD Controller}
\ 

\begin{figure}[thpb]
  \centering
  \includegraphics[scale=0.5]{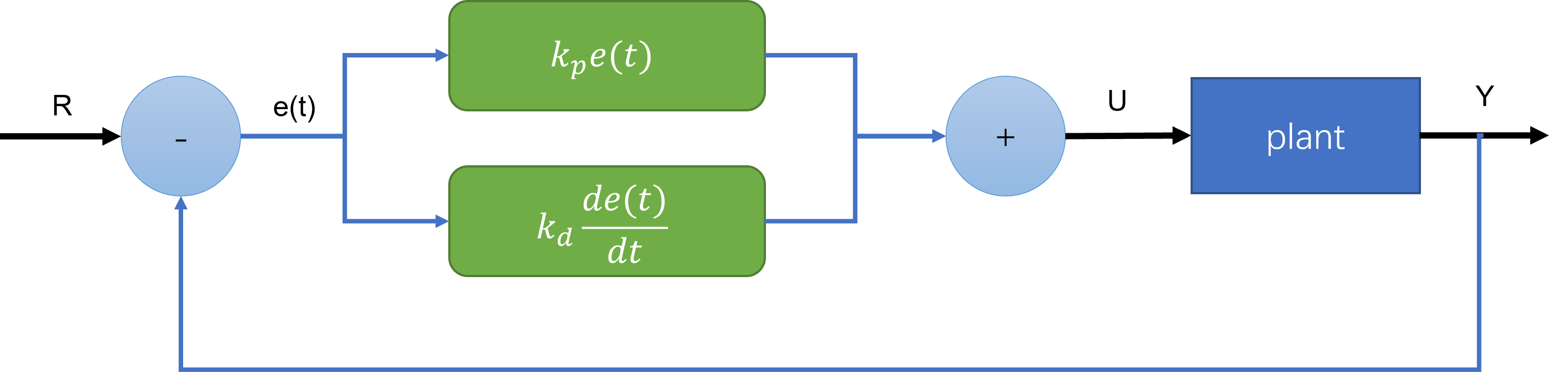}
  \caption{Conceptual structure of conventional PD controller}
  \label{PD-structure}
\end{figure}

Fig.\ref{PD-structure} is the conceptual structure of a conventional PD controller. $R$ is the reference signal, $e(t)$ is the error in time $t$, $U$ is the control signal and $Y$ is the output. The core modules of PD controller are proportional and derivative module, and the mathematical expression is
\begin{equation}
    U=K_pe(t)+k_d\frac{de(t)}{dt},
\end{equation}
where $k_p,k_d$ are the proportional gain and derivative gain. PD controller is a simplified version of Proportional-Integral-Derivative (PID) controller, which is widely adopted in the industry.

\subsubsection{Type-1 Fuzzy Logic System}
\ 

\begin{figure}[thpb]
  \centering
  \includegraphics[scale=0.5]{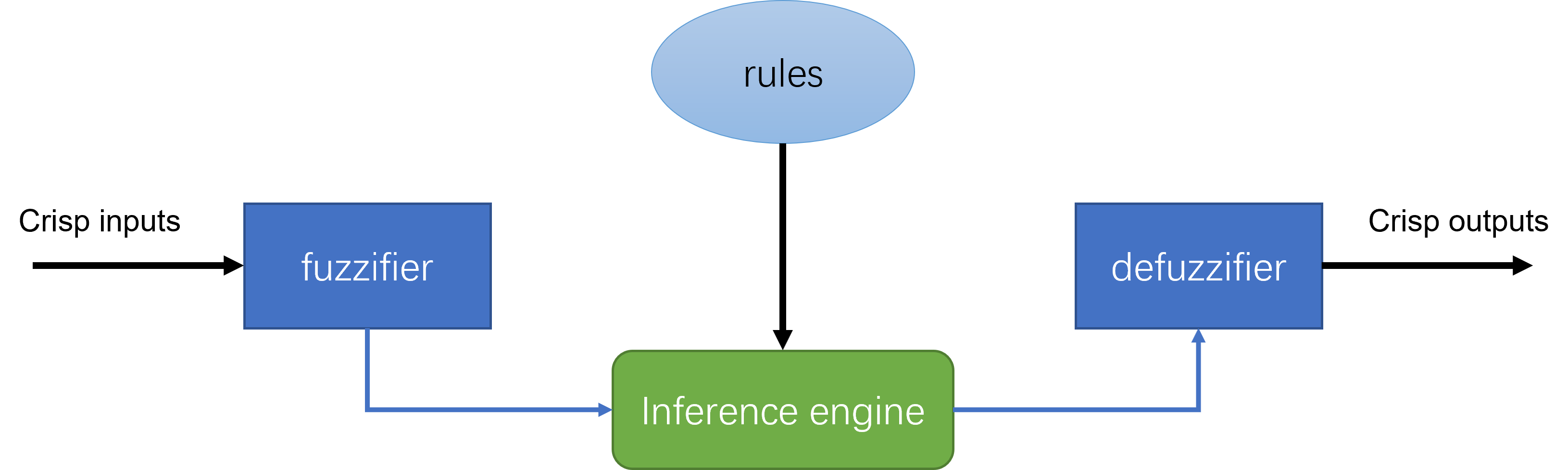}
  \caption{General workflow of type-1 fuzzy system}
  \label{type1fuzzy}
\end{figure}

Fig.\ref{type1fuzzy} presents the general workflow of type-1 fuzzy logic systems. The crisp inputs are first fuzzified by the fuzzifier, mapping to a value between 0 and 1 using membership functions (MF). Then it will be processed by the rule-based inference engine. Lastly, the type-1 output fuzzy set will be defuzzified to retrieve a crisp outputs.

Among them, the inference engine is the core component of fuzzy system. The IF-THEN rules of a multi-input-multi-output (MIMO) fuzzy systems are in the following format:
\begin{equation}
    \text{RULE i: IF } x_1(t) \text{ is } M^i_1 \text{ AND ... AND }x_n(t) \text{ is } M^i_n \text{ THEN } y_1^i(t) \text{ AND ... AND }y_m^i(t),
\end{equation}
where $x_i,i=1,2,...,n$ is the linguistic input variables, $y_j^i(t),j=1,2,...,m$ is the intermediate output of the output $y_j(x(t))$ under $i^{th}$ rule. $M^i_j$ is the fuzzy term of the $j^{th}$ linguistic variable under $i^{th}$ rule. In Sugeno fuzzy inference system, the final output is 
\begin{equation}
    y_j(x(t))=\sum_{i=1}^\mathbb{N} \omega_i(X(t)) y_j^i(t).
\end{equation}
$X(t)=[x_1(t),x_2(t),...,x_N(t)]$, and $\mathbb{N}$ is the total number of fuzzy rules. $\omega_i(X(t))$ is the normalized firing strength of the $i^{th}$ rule:
\begin{equation}
    \omega_i(X(t))=\frac{\prod_{k=1}^n \mu_{M^i_k}(X(t))}{\sum_{m=1}^\mathbb{N}{\prod_{k=1}^n \mu_{M^m_k}(X(t))} }.
\end{equation}
Also, $\omega_i(X(t))>0, \forall i$, and $\sum \omega_i(X(t))=1$. $\mu_{M^i_k}(X(t))$ is the grade of the MF of fuzzy term $M^i_k$.

\subsubsection{Fuzzy PD Controller}
\ 

\begin{figure}[thpb]
  \centering
  \includegraphics[scale=0.4]{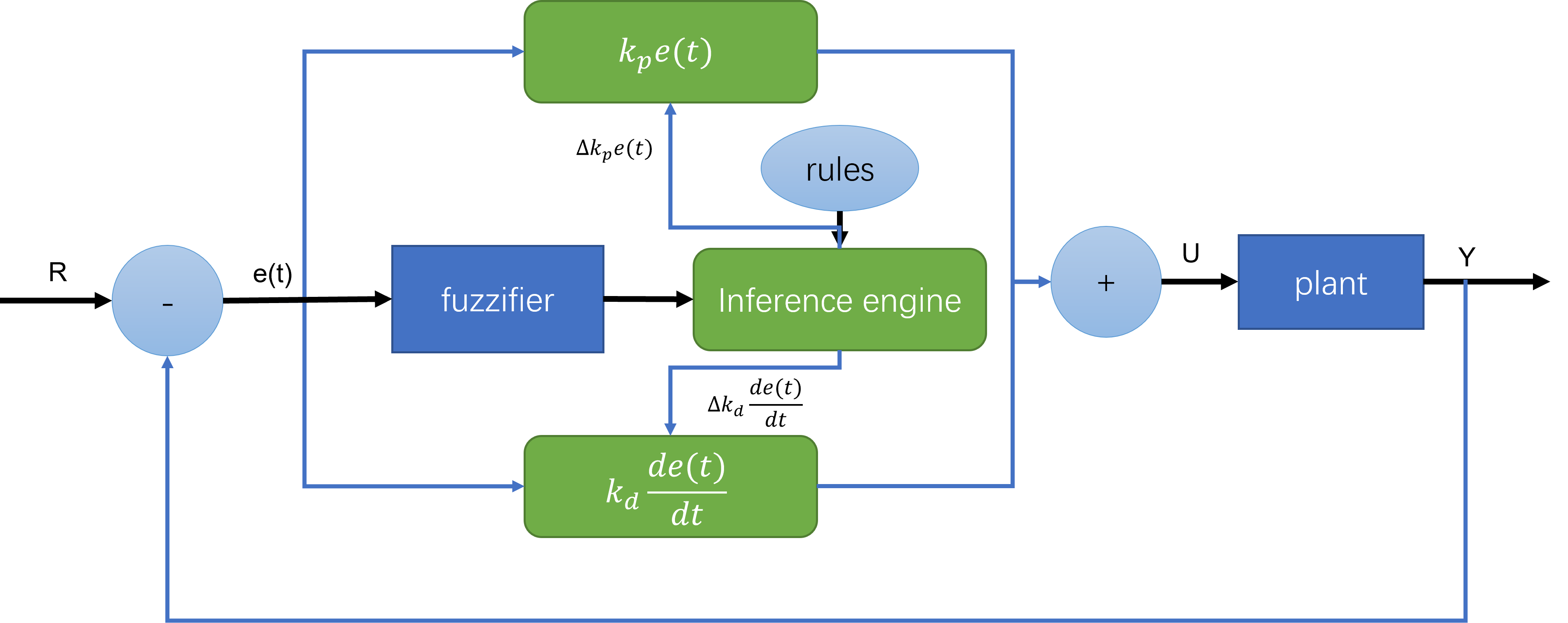}
  \caption{Fuzzy PD structure}
  \label{fuzzypd}
\end{figure}

Fig.\ref{fuzzypd} is the structure of single fuzzy PD controller. The fuzzy inference system is used to adjust the proportional and derivative gains of conventional PD controller.

\subsection{Bayesian Optimization}
Bayesian Optimization (BO) is a powerful algorithm in combinatorial optimization realm, especially when the original cost function is expensive to measure. Among all the other optimization algorithms, BO stands out in terms of its ability to search for global optimality and data efficiency. To do that, BO maintains a cheap surrogate model that replaces the real expensive cost function. The most commonly used surrogate model is Gaussian Regression, which integrates all sampled data together to form a global probabilistic model that describes its belief on the cost value distribution. With the global model, BO calculates the next location to sample with the highest probability to retrieve the best result. Compared with other popular optimization algorithms, BO is free from gradient information, and can find relatively satisfying result quickly \cite{NeumannBrosig2020DataEfficientAW}. 

The simplest form of BO is Sequential Model-based Optimization \cite{Snoek2012PracticalBO}, of which the pseudo code is displayed in Alg.\ref{BO}. Firstly, the surrogate model $\mathcal{M}$, cost function $f$, acquisition function $S$ and parameter domain $\mathcal{X}$ are initialized. Then we sample randomly from $\mathcal{X}$ to retrieve an initial data base $\mathcal{D}$. Then for a fixed number of steps $T$, $\mathcal{M}$ is first updated as posterior distribution $p(y|x,D)$ to best represent all the data inside $\mathcal{D}$. Then the next location $x_i$ to be explored is retrieved by maximizing acquisition function $\mathbb{S}$, which is some form of evaluation on the probability of getting lower cost. Then the real cost $f(x_i)$ is evaluated and added to the database $\mathcal{D}$. In this paper, the acquisition function is selected as Upper Confidence Bound (UCB) \cite{Snoek2012PracticalBO}, represented as 
\begin{equation}
    UCB=\mu(x_i)+h\sigma(x_i).
\end{equation}
$\mu(x_i)$ is the mean value, and $\sigma(x_i)$ is the covariance. UCB can be simply interpreted as an upper bound of our confidence on certain location. The coefficient $h$ adjusts the exploration extent, and is selected as $2.576$ \cite{BOOO}.

\begin{algorithm}[H]
  \caption{Sequential Model-based Optimization} 
  \begin{algorithmic}[1]
    \State \textbf{Input:} f,$\mathcal{X}$,$\mathcal{S}$,$\mathcal{M}$
    \State $\mathcal{D}$ $\leftarrow$ INITSAMPLES(f,$\mathcal{X}$) 
    \For {$i \leftarrow |\mathcal{D}| \ to \ T$}
        \State $p(y|x,\mathcal{D}) \leftarrow  FITMODEL(\mathcal{M},\mathcal{D})$
        \State $x_i \leftarrow argmax_{x \in \mathcal{X}}\mathcal{S}(x,p(y|x,\mathcal{D}))$
        \State $y_i \leftarrow  f(x_i)$
        \State $\mathcal{D} \leftarrow  \mathcal{D} \cup (x_i,y_i)$
    \EndFor
  \end{algorithmic} \label{BO}
\end{algorithm}

In this paper, BO is implemented in all tuning processes, including for PD controllers and fuzzy logic regulators (FLC). The considerations behind are that testing real systems with un-verified sets of parameters is unsafe, which may damage the mechanics and bring danger to human operators. Besides, fewer times of runs means less wear and tear of the system \cite{Knig2021SafeAE}. Therefore, BO is suitable to be integrated into the design procedures of our proposed algorithms.

\section{Controller Design and Analysis}
This section articulates the controller design procedures. Firstly, a conceptual graph of the controller is presented, with an introduction to the core idea. Secondly, cascaded PD controller is designed. After that, stability analysis of the cascaded PD controller is implemented through transfer function and Jacobian matrix. Lastly, the design of fuzzy logic regulator is specified.

\subsection{Framework Overview} \label{FrameworkOverview}
\begin{figure}[thpb]
  \centering
  \includegraphics[scale=0.5]{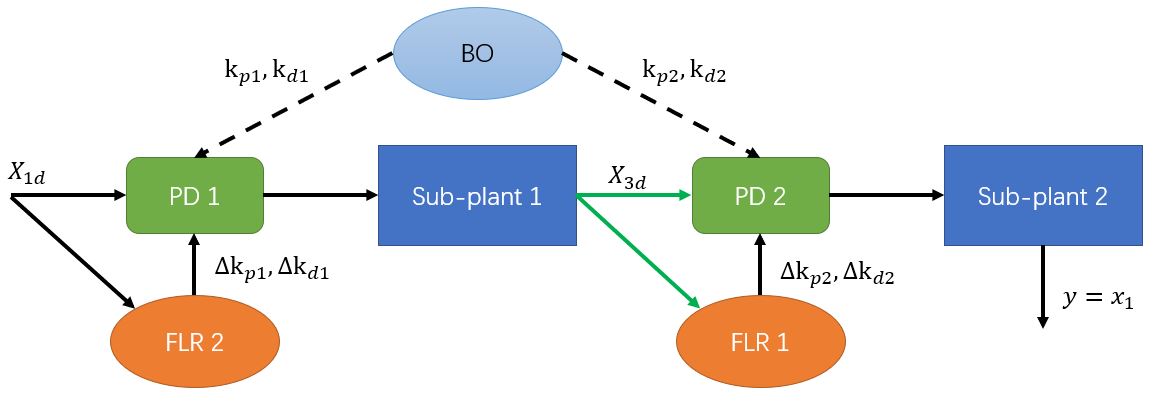}
  \caption{Framework of controller design}
  \label{framework}
\end{figure}

Fig.\ref{framework} illustrates the conceptual framework of the controllers. Inside our designed framework, the original fourth-order system (\ref{dynamicmodel3}) is divided into two second-order sub-plants, which are expressed in (\ref{subplant1}) and (\ref{subplant2}). The inspiration of cascaded PD controller for the system is to use one PD controller each for those two sub-plants, which are second-order systems. It is expected that if two sub-plants can be stabilized separately under two separate PD controllers, the original system can be stable. However, the coupling term $x_3$ that appeared in (\ref{subplant1}) should be handled properly, which will be explained further in Sec.\ref{CascadedPDControllerDesign}.

\begin{equation}
sub-plant 1:
\begin{cases}
\dot{x}_1 = x_2 \\
\dot{x}_2 = -\frac{mgl}{I_l}cosx_1-\frac{k}{I_l}(x_1-x_3)+d_1 \\
\end{cases} \label{subplant1}
\end{equation}

\begin{equation}
sub-plant 2:
\begin{cases}
\dot{x}_3 = x_4 \\
\dot{x}_4 = \frac{k}{I_m}(x_1-x_3)-\frac{\mu}{I_m}x_4 + \frac{1}{I_m}u+d_2 \\
\end{cases} \label{subplant2}
\end{equation}

The workflow of our controller is specified in the following. Firstly, the reference signal $X_{1d}=[x_{1d},\dot{x}_{1d}]^T$ is input into the first controller PD1, where the desired intermediate torque for the first second-order system is computed. Through combining  (\ref{subplant1}), a referenced signal $X_{3d}=[x_{3d},\dot{x}_{3d}]$ for sub-plant2 is derived. This will be input into the second PD controller (namely PD2) to stabilize the sub-plant2. Upon all that, BO is implemented during the tuning process to render appropriate parameters $k_{p1},k_{d1}$ and $k_{p2},k_{d2}$ for PD1 and PD2 respectively. $k_{p1},k_{p2}$ represent the proportional gains, and $k_{d1},k_{d2}$ are the derivative gains. After that, two fuzzy logic regulators (FLR1 and FLR2) will determine online the regulation values $\Delta k_{p1}, \Delta k_{d1},\Delta k_{p2},\Delta k_{d2}$ for the parameters of the PD controllers.

\subsection{Cascaded PD Controller Design} \label{CascadedPDControllerDesign}
In this section, the design process of cascaded PD controllers is elaborated, Firstly, PD1 output is transferred to reference signal $x_{3d}$. Secondly, $x_{3d}$ is utilized to design the second PD controller. Lastly, certain compensation and simplification is made to transfer sub-plant2 into a standard second-order serial integrator with disturbance.

Consider a serial integrator with disturbance $d_1$, being controlled by a PD controller with proper parameters in (\ref{dynamics1}), where $x_{1d},\dot{x}_{1d}$ are the given reference signal. 

\begin{equation}
\begin{cases}
\dot{x}_1 = x_2 \\
\dot{x}_2 = u_{PD1} + d_1 \\
u_{PD1}=k_{p1} (x_{1d}-x_1) + k_{d1} (\dot{x}_{1d}-\dot{x}_1)
\end{cases}
\label{dynamics1}
\end{equation}

Associating with (\ref{subplant1}), it can be expected that if $-\frac{mgl}{I_l}cosx_1-\frac{k}{I_l}(x_1-x_3)=u_{PD1}$, then (\ref{subplant1}) can be stabilized. While $x_1,x_2$ are the state variables of (\ref{subplant1}), $x_3$ is an external variable which can be utilized freely for controller design. Therefore, we assign the desired $x_3$ to be $x_{3d}$ that satisfies 
\begin{equation}
    -\frac{mgl}{I_l}cosx_1-\frac{k}{I_l}(x_1-x_{3d})=u_{PD1}.
    \label{PD1}
\end{equation}
Therefore, the reference signal for the motor is derived as 
\begin{equation}
    x_{3d}=\frac{u_{PD1}I_l}{k}+x_1+\frac{mglcosx_1}{k}.
\end{equation}

Consequently, the direct PD controller for (\ref{subplant2}) becomes apparent by assigning $u=u_{PD2}$, which can be expressed as 
\begin{equation}
\begin{cases}
\dot{x}_3 = x_4 \\
\dot{x}_4 = \frac{k}{I_m}(x_1-x_3)-\frac{\mu}{I_m}x_4 + \frac{1}{I_m}u_{PD2}+d_2 \\
u_{PD2}=k_{p2} (x_{3d}-x_3) + k_{d2} (\dot{x}_{3d}-\dot{x}_3)
\end{cases} 
\label{dynamics2}
\end{equation}

Further, to compensate for the term $\frac{k}{I_m}(x_1-x_3)$ that illustrates coupling with sub-plant1, $u$ is modified as 
\begin{equation}
    u_2=u_{PD2}+u_{PD1}I_l+mglcosx_1. \label{u2}
\end{equation}
Integrating (\ref{u2}) into (\ref{subplant2}) and (\ref{PD1}), the resulting dynamics is shown in (\ref{dynamicmodel3}). 
\begin{equation}
\begin{cases}
\dot{x}_3 = x_4 \\
\dot{x}_4 = u_{2} + d_2' \\
u_{2}=\frac{k_{p2}+k}{I_m}(x_{3d}-x_3) + \frac{k_{d2}}{I_m} (\dot{x}_{3d}-\dot{x}_3)\\
d_2'=-\frac{\mu}{I_m}x_4 + d_2
\end{cases} 
\label{dynamics3}
\end{equation}

However, the derivative reference signal $\dot{x}_{3d}$ is not given directly and should be calculated as 
\begin{equation}
    \dot{x}_{3d} = \frac{d(\frac{u_{PD1}I_l}{k}+x_1+\frac{mglcosx_1}{k})}{dt}=\frac{d(\frac{u_{PD1}I_l}{k})}{dt}+\dot{x}_1-\frac{mglsinx_1}{k}\dot{x}_1. \label{dynamics4}
\end{equation}
This calculation is possible but very complicated, especially when it involves the derivative of PD1 controller. For simplicity in this paper, we assign \begin{equation}
    \dot{x}_{3d}=0. \label{x3dd}
\end{equation}

\textbf{Observation 1}: If sub-plant1 and sub-plant2 can be stabilized separately, we would expect the whole system to be stable. Indeed, it can be proved that system (\ref{PD1}) and (\ref{dynamics4}) can be stabilized separately\cite{Zhao2017PIDCD}. However, a joint analysis is still required to ensure stability of the fourth-order system, which will be detailed in Sec.\ref{StabilityAnalysisUsingJacobianMatrix}.

\textbf{Observation 2}: The cascaded PD controller in this paper is different from the conventional one. Conventional cascaded PD controller works in adjacent order of the system. For example, one PD controller to assign desired velocity, and the other PD controller to control the acceleration \cite{Andrade2021UnmannedAV}. In contrast, the PD1 controller in this paper serves as the acceleration controller for sub-plant1, as well as the calculator of the reference signal for sub-plant2. And PD2 controller is the acceleration controller for sub-plant2.

\subsection{Type-1 Fuzzy Logic Regulator Design}
\begin{figure}[htbp]
\centering
\begin{minipage}[t]{0.48\textwidth}
\centering
\includegraphics[scale=0.48]{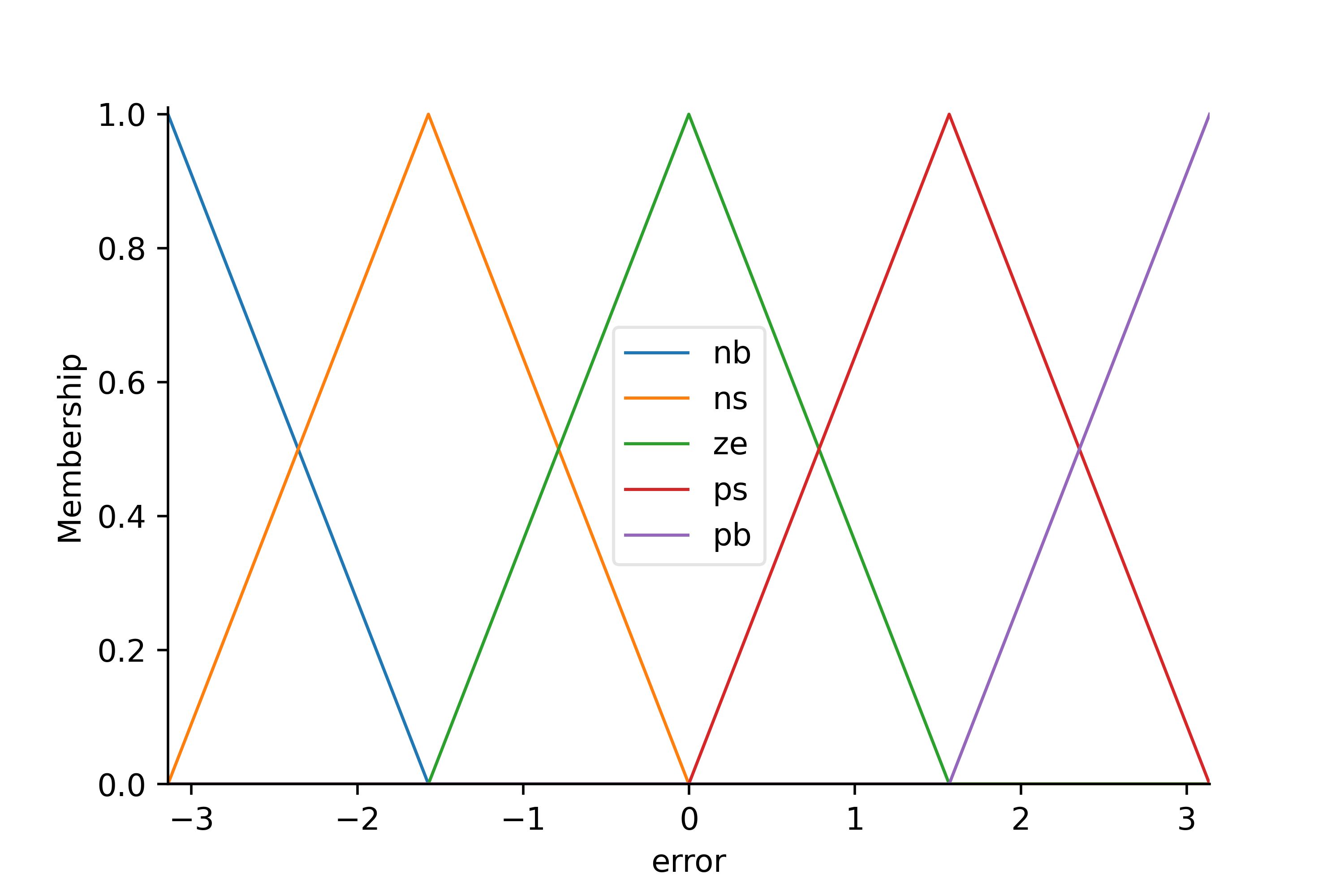}
\caption{Membership function of error} \label{errormf}
\end{minipage}
\begin{minipage}[t]{0.48\textwidth}
\centering
\includegraphics[scale=0.48]{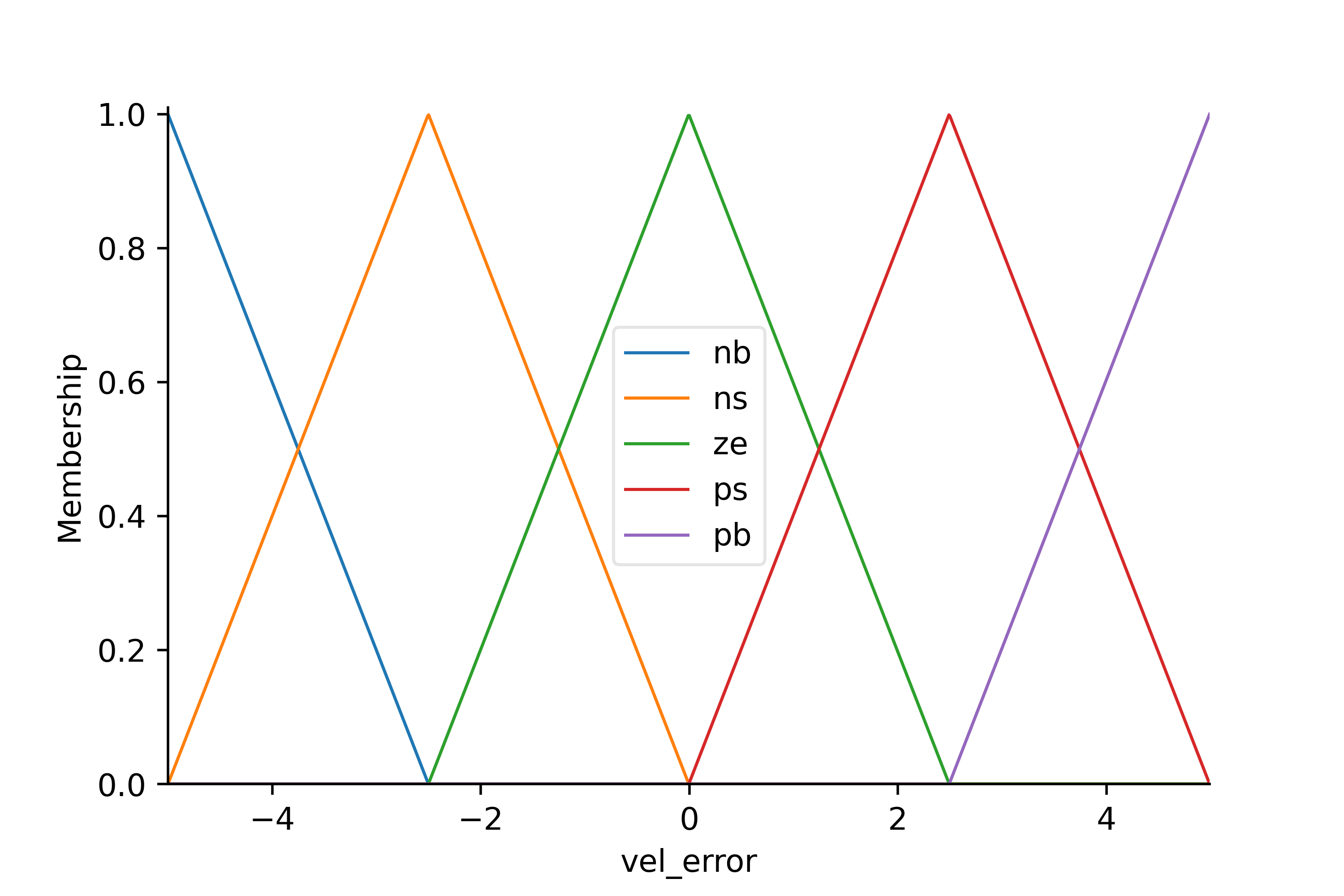}
\caption{Membership function of velocity error} \label{velerrormf}
\end{minipage}
\end{figure}

The fuzzy logic regulator (FLR) is used to adjust the parameters of PD controller adaptively. Two FLRs are required, with each deals with one PD controller. For the first FLR, the inputs are $e_1, e_2$, and the outputs are $\Delta k_{p1},\Delta k_{d1}$. For the second FLR, the inputs are $e_3, e_4$, and the outputs are $\Delta k_{p2},\Delta k_{d2}$. The inputs will pass through a fuzzification module, and then will be processed by fuzzy inference module using predefined fuzzy rules. At last, a crisp value is output using defuzzification module. All the inputs and outputs are described by 5 linguistic variables, namely Negative Big (NB), Negative Small (NS), Zero (ZE), Positive Small (PS) and Positve Big (PB). The memberships functions are selected as triangular membership functions, and are divided evenly that spread across the domain of variables. The inputs of the FLRs are manually set as 
\begin{equation}
    e_1,e_3 \in [-\pi,\pi]rad;e_2,e_4 \in [-5,5]rad/s.
\end{equation}
The membership function of the inputs are depicted in Fig.\ref{errormf} and Fig.\ref{velerrormf}. Similarly, the domain of the outputs are defined using unknown parameters below, which are to be tuned by BO.
\begin{equation}
    \Delta k_{p1} \in [\Delta k_{p1}^l, \Delta k_{p1}^u] \label{anticident1}
\end{equation}
\begin{equation}
    \Delta k_{d1} \in [\Delta k_{d1}^l, \Delta k_{d1}^u] \label{anticident2}
\end{equation}
\begin{equation}
    \Delta k_{p2} \in [\Delta k_{p2}^l, \Delta k_{p2}^u] \label{anticident3}
\end{equation}
\begin{equation}
    \Delta k_{d2} \ in [\Delta k_{d2}^l, \Delta k_{d2}^u] \label{anticident4}
\end{equation}

The fuzzy rules are the key element to fuzzy inference module. The fuzzy rules for our PD controllers are defined in Tab.\ref{k_prule} and Tab.\ref{k_drule}, in which $e$ represents the first input, namely the angular error in our experiments. $de$ is the second input, which is the angular velocity error in this paper. The overall notion of fuzzy rule design is that when the errors are big, proportional gain should be increased to compensate for it, while derivative gain should be decreased. When the errors are small, the proportional gain should be decreased and the derivative gain should be decreased to prevent overshoot.

\begin{table}[!htbp]
\centering
\caption{Rule table for $\Delta k_p$}
\begin{tabular}{cccccc}
\hline
e/ de & NB & NS & ZE & PS & PB\\
\hline
NB & NB & NB & NS & NS & ZE\\
NS & NB & NS & NS & ZE & PS\\
ZE & NS & NS & ZE & PS & PS\\
PS & NS & ZE & PS & PS & PB\\
PB & ZE & PS & PS & PB & PB\\

\hline \label{k_prule}
\end{tabular}
\end{table}
\begin{table}[!htbp]
\centering
\caption{Rule table for $\Delta k_d$}
\begin{tabular}{cccccc}
\hline
e/ de & NB & NS & ZE & PS & PB\\
\hline
NB & PB & PB & PS & PS & ZE\\
NS & PB & PS & PS & ZE & NS\\
ZE & PS & PS & ZE & NS & NS\\
PS & PS & ZE & NS & NS & NB\\
PB & ZE & NS & NS & NB & NB\\

\hline \label{k_drule}
\end{tabular}
\end{table}

\subsection{Simplified Linear System Transfer Function Analysis} \label{SimplifiedLinearSystemTransferFunctionAnalysis}
In this section, the system performance without FLRs is analyzed using transfer function. Through calculating the poles of the characteristic function of the resulting system, the stability analysis can be carried out. To do that, a nominal model with $g=0$ is used in this section, which means the model is a simplified linear version of (\ref{dynamicmodel3}) without disturbance. With that being said, it can still render a good estimate of the original system dynamics, or even becomes the real system analysis if all nonlinear terms and unknown disturbance are properly compensated. 

Combining (\ref{dynamicmodel3})(\ref{PD1})(\ref{u2})(\ref{x3dd}), and setting $d_1=0,d_2=0,g=0$, the whole system dynamics is 
\begin{equation}
\begin{cases}
\dot{x}_1 = x_2 \\
\dot{x}_2 = -\frac{k}{I_l}(x_1-x_3) \\
\dot{x}_3 = x_4 \\
\dot{x}_4 = \frac{k_{p1}(k_{p2}+k)I_l}{kI_m}(x_{1d}-x_1) +\frac{k_{d1}k_{p1}I_l}{kI_m}(\dot{x}_{1d}-x_2)+\frac{(k_{p2}+k)(x_1-x_3)}{I_m} - \frac{k_{d2}x_4}{I_m} - \frac{\mu x_4}{I_m} \\

\end{cases} \label{dynamicmodel4}
\end{equation}

Assuming the initial values of all state variables to be 0. Taking the Laplace transformation of $(\dot{x}_1,\dot{x}_2)$ from (\ref{dynamicmodel4}):
\begin{equation}
    s^2 X_1(s)=-\frac{k}{I_l}(X_1(s)-X_3(s)), \label{lap12}
\end{equation}
where $s$ is the Laplace variable, and $X_i(s), i=1,2,3,4$ are the Laplace transformation results of corresponding variables $x_i$. For simplicity, the dependent variable $s$ will be omitted, and $X_i(s)$ will be written as $X_i$. From (\ref{lap12}), the following transfer function is derived:
\begin{equation}
    \frac{X_1}{X_3}=\frac{k}{I_ls^2+k}. \label{tf13}
\end{equation}

Similarly, taking the Laplace transform of $(\dot{x}_3,\dot{x}_4)$ from (\ref{dynamicmodel4}):

\begin{equation}
    s^2X_3=\frac{k_{p1}(k_{p2}+k)I_l}{kI_m}(X_{1d}-X_1) +\frac{k_{d1}k_{p1}I_l}{kI_m}s(X_{1d}-X_1)+\frac{(k_{p2}+k)(X_1-X_3)}{I_m} - \frac{k_{d2}sX_3}{I_m} - \frac{\mu sX_3}{I_m}
\end{equation}

Merging similar items and move $X_3$ all to the left:
\begin{equation}
    X_3=\frac{\frac{k_{p1}(k_{p2}+k+sk_{d1})I_l}{k}(X_{1d}-X_1) +(k_{p2}+k)X_1}{I_ms^2+(K_{d2}+\mu)s+(k_{p2}+k)} \label{tf3}
\end{equation}
Multiplying (\ref{tf13}) by (\ref{tf3}):
\begin{equation}
    X_1=\frac{k_{p1}(k_{p2}+k+sk_{d1})I_l(X_{1d}-X_1) +k(k_{p2}+k)X_1}{[I_ms^2+(K_{d2}+\mu)s+(k_{p2}+k)](I_ls^2+k)} \label{tf1}
\end{equation}

Rearranging (\ref{tf1}), and then the transfer function 
\begin{equation}
    \frac{X_1}{X_{1d}}=\frac{k_{p1}(k_{p2}+k+sk_{d1})}{I_mI_ls^4+(I_lK_{d2}+I_l\mu)s^3+[I_mk+I_l(k_{p2}+k)]s^2+(k\mu + kk_{d2} +k_{p1}k_{d1})s+k_{p1}(k_{p2}+k)} \label{tf11d}
\end{equation}

(\ref{tf11d}) depicts the response of $x_1$ given the reference signal $x_{1d}$. It is obvious that the denominator has $4^{th}$ order, and the stability is determined by the poles of the transfer function (\ref{tf11d}). Since it is determined by all the parameters of the system, it will be calculated numerically in Sec.\ref{StabilityAnalysisofCascadedPDController}.

\subsection{Stability Analysis Using Jacobian Matrix} \label{StabilityAnalysisUsingJacobianMatrix}
In Sec.\ref{SimplifiedLinearSystemTransferFunctionAnalysis}, transfer function analysis is implemented on a simplified linear model. Although the system turns out to be stable, it neglects the nonlinearity and disturbance. Therefore, this section proves that our proposed controller is asymptotically stable even with nonlinearity and disturbance, given that the disturbance satisfies certain conditions. Zhao et al. has proved the global asymptotic stability of a general uncertain $2^{nd}$ order dynamic system can be achieved given certain assumptions \cite{Zhao2017PIDCD}. Our analysis here is an extension of theirs from $2^{nd}$ order to $4^{th}$ order dynamics.

Firstly, the error equations are defined:
\begin{equation}
\begin{cases}
e_1 = x_{1d}-x_1 \\
e_2 = \dot{x}_{1d}-\dot{x}_1=\dot{x}_{1d}-x_2 \\
e_3 = x_{3d} - x_3 = \frac{u_{PD1}I_l}{k}+x_1+\frac{mglcosx_1}{k} - x_3 \\
e_4 = \dot{x}_{3d}-\dot{x}_3=\dot{x}_{3d}-x_4

\end{cases} \label{errors1}
\end{equation}

Integrating (\ref{PD1})(\ref{x3dd}) into (\ref{errors1}):
\begin{equation}
\begin{cases}
e_1 = x_{1d}-x_1 \\
e_2 = \dot{x}_{1d}-x_2 \\
e_3 = \frac{(k_{p1}e_1+k_{d1}e_2)I_l}{k}+x_1+\frac{mglcosx_1}{k} - x_3 \\
e_4 = -x_4

\end{cases} \label{errors2}
\end{equation}

\textbf{Definition 1}: For a general second-order dynamic system with disturbance 
\begin{equation}
\begin{cases}
\dot{x}_1 = x_2 \\
\dot{x}_2 = f(x_1,x_2,t) + u,\\
\end{cases} \label{generaldynamicmodel}
\end{equation}
a special functional space is defined as follow:
\begin{equation}
\mathcal{F}_{L_1,L_2}(x_1, x_2)=\Big\{ f \in C^1(\mathbb{R}^2 \times \mathbb{R}^+) \Big| \big|\frac{\partial f}{\partial x_1} \big| \leq L_1,  \big|\frac{\partial f}{\partial x_2} \big| \leq L_2, \forall x_1,x_2 \in \mathbb{R}, \forall t \in \mathbb{R}^+ \Big\},
\end{equation}
where $L_1,L_2$ are positive constants, and $C^1(\mathbb{R}^2 \times \mathbb{R}^+)$ denotes the functional space mapping $\mathbb{R}^2 \times \mathbb{R}^+$ to $\mathbb{R}$, which are locally Lipschitz continuous in $(x_1,x_2)$ uniformly in $t$, and piecewise continuous in $t$.

\textbf{Assumption 1}: There exists positive constants $L_{11},L_{12},L_{21},L_{22}$ that satisfy the following:
\begin{equation}
    d_1 \in \mathcal{F}_{L_{11},L_{12}}(x_1, x_2)
\end{equation}
\begin{equation}
    d_2 \in \mathcal{F}_{L_{21},L_{22}}(x_3, x_4)
\end{equation}

\textbf{Theorem 1}: For a class of $4^{th}$-order dynamic systems (\ref{dynamicmodel3}), using controller presented in (\ref{u2}) and (\ref{x3dd}), the system can achieve locally asymptotic stability under Assumption 1, if the followings are satisfied:
\begin{equation}
    k_{d1}>L_{12},k_{p1}>L_{11},\frac{\mu+k_{d2}}{I_m}>L_{22}, \frac{k+k_{p2}}{I_m}>L_{21} \label{stabilitycondition}
\end{equation}

\indent \textbf{Proof}. 
Taking the derivative of (\ref{errors2}) and integrating into (\ref{dynamicmodel3}):

\begin{equation}
\begin{cases}
\dot{e}_1 = e_2 \\
\dot{e}_2 = \ddot{x}_{1d}+\frac{mgl}{I_l}cosx_1+\frac{k}{I_l}(x_1-x_3)-d_1 \\
\dot{e}_3 = e_4 \\
\dot{e}_4 = -\dot{x}_4 = \frac{\mu}{I_m}x_4-\frac{(k_{p2}+k)e_3+k_{d2}e_4}{I_m} - d_2

\end{cases} \label{errors_d1}
\end{equation}

To remove all state variables $\{x_i,i=1,2,3,4 \}$ in (\ref{errors_d1}), integrate it with (\ref{errors2}), and we have the dynamics of the error vectors. 
\begin{equation}
\begin{cases}
\dot{e}_1 = e_2 \\
\dot{e}_2 = \ddot{x}_{1d}-k_{p1}e_1-k_{d1}e_2+\frac{k}{I_l}e_3-d_1 \\
\dot{e}_3 = e_4 \\
\dot{e}_4 = -\frac{\mu+k_{d2}}{I_m}e_4-\frac{k+k_{p2}}{I_m}e_3-d_2
\end{cases} \label{errors_d2}
\end{equation}

Denote the vector field of (\ref{errors_d2}) as $F(e_1,e_2,e_3,e_4)$, i.e. 
\begin{equation}
F(e_1,e_2,e_3,e_4)=
\left [
\begin{array}{c}
    e_2 \\
    \ddot{x}_{1d}-k_{p1}e_1-k_{d1}e_2+\frac{k}{I_l}e_3-d_1 \\
    e_4 \\
    -\frac{\mu+k_{d2}}{I_m}e_4-\frac{k+k_{p2}}{I_m}e_3-d_2
\end{array}
\right ]
\end{equation}

Then the Jacobian matrix of $F(e_1,e_2,e_3,e_4)$ is 
\begin{equation}
DF(e_1,e_2,e_3,e_4)=
\left [
\begin{array}{cccc}
    0 & 1 & 0 & 0 \\
    \frac{\partial \ddot{x}_{1d}}{\partial e_1}-\frac{\partial d_1}{\partial e_1}-k_{p1} & 
    \frac{\partial \ddot{x}_{1d}}{\partial e_2}-\frac{\partial d_1}{\partial e_2}-k_{d1} & 
    \frac{\partial \ddot{x}_{1d}}{\partial e_3} + \frac{k}{I_l}& 
    \frac{\partial \ddot{x}_{1d}}{\partial e_4} \\
    0 & 0 & 0 & 1 \\
    0 & 
    0 & 
    -\frac{k+k_{p2}}{I_m}-\frac{\partial d_2}{\partial e_3} & 
    -\frac{\mu+k_{d2}}{I_m}-\frac{\partial d_2}{\partial e_4}
\end{array}
\right ] \label{DF}
\end{equation}

Usually, the reference signal is not dependent on the state variables, but only on time $t$. Therefore, (\ref{DF}) can be simplified to 
\begin{equation}
DF(e_1,e_2,e_3,e_4)=
\left [
\begin{array}{cccc}
    0 & 1 & 0 & 0 \\
    -\frac{\partial d_1}{\partial e_1}-k_{p1} & 
    -\frac{\partial d_1}{\partial e_2}-k_{d1} & 
    \frac{k}{I_l}& 
    0 \\
    0 & 0 & 0 & 1 \\
    0 & 
    0 & 
    -\frac{k+k_{p2}}{I_m}-\frac{\partial d_2}{\partial e_3} & 
    -\frac{\mu+k_{d2}}{I_m}-\frac{\partial d_2}{\partial e_4}
\end{array}
\right ] \label{DF1}
\end{equation}

The eigenvalues of (\ref{DF1}) have closed-form solutions:
\begin{equation}
    \lambda_1=-\frac{1}{2}\frac{\partial d_1}{\partial e_2}-\frac{1}{2}k_{d1}- \frac{1}{2}\sqrt{ \Big[(\frac{\partial d_1}{\partial e_2}+k_{d1})^2-4(\frac{\partial d_1}{\partial e_1}+k_{p1}) \Big]}
\end{equation}
\begin{equation}
    \lambda_2=-\frac{1}{2}\frac{\partial d_1}{\partial e_2}-\frac{1}{2}k_{d1} + \frac{1}{2}\sqrt{ \Big[(\frac{\partial d_1}{\partial e_2}+k_{d1})^2-4(\frac{\partial d_1}{\partial e_1}+k_{p1}) \Big]}
\end{equation}
\begin{equation}
    \lambda_3=-\frac{1}{2}\frac{\mu+k_{d2}}{I_m}-\frac{1}{2}\frac{\partial d_2}{\partial e_4} - \frac{1}{2}\sqrt{ \Big[(\frac{\mu+k_{d2}}{I_m}+\frac{\partial d_2}{\partial e_4})^2-4(\frac{k+k_{p2}}{I_m}+\frac{\partial d_2}{\partial e_3}) \Big]}
\end{equation}
\begin{equation}
    \lambda_4=-\frac{1}{2}\frac{\mu+k_{d2}}{I_m}-\frac{1}{2}\frac{\partial d_2}{\partial e_4} + \frac{1}{2}\sqrt{ \Big[(\frac{\mu+k_{d2}}{I_m}+\frac{\partial d_2}{\partial e_4})^2-4(\frac{k+k_{p2}}{I_m}+\frac{\partial d_2}{\partial e_3}) \Big]}
\end{equation}
If (\ref{stabilitycondition}) and Assumption 1 are satisfied, all four eigenvalues have negative real parts. Note that $[e_1,e_2,e_3,e_4]^T=[0,0,0,0]^T$ is obviously the set point of (\ref{errors_d2}). Therefore, the system is asymptotically stable \cite{Bhatia1970StabilityTO}, converging to $[e_1,e_2,e_3,e_4]^T=[0,0,0,0]^T$. In other words, all orbits starting close enough to the set point tends asymptotically to it.

\textbf{Observation 3}: It is tempting to extend the conclusion into globally asymptotic stability according to Markus-Yamabe’s theorem \cite{Feler1995APO}. Nevertheless, Markus-Yamabe’s theorem currently only holds for $2^{nd}$-order systems, and counterexamples have been investigated in higher-order systems \cite{KUZNETSOV2018138}. As for what is the extreme of initial points to ensure asymptotic stability, we can implement numerical experiments to determine. Nevertheless, one interesting fact about (\ref{DF1}) is that the existence of the coupling term $\frac{k}{I_l}$ does not affect the result of eigenvalues. Namely, the stability condition of this coupled system is the same as that if the coupling between the sub-plant1 and sub-plant2 disappears and that they are totally decoupled.

Further, the controller with FLR integrated can be analyzed under the same assumptions and conditions. Due to the rationale of fuzzy logic systems, the outputs of a FLR are limited by the lower and upper bounds of the antecedents as shown in (\ref{anticident1})-(\ref{anticident4}). Therefore, the stability condition should be specified as 
\begin{equation}
    k'_{d1}+\Delta k_{d1}^l>L_{12} \label{fuzzycondition1}
\end{equation}
\begin{equation}
    k'_{p1}+\Delta k_{p1}^l>L_{11} \label{fuzzycondition2}
\end{equation}
\begin{equation}
    \frac{\mu+k_{d2}+\Delta k_{d2}^l}{I_m}>L_{22} \label{fuzzycondition3}
\end{equation}
\begin{equation}
    \frac{k+k_{p2}+\Delta k_{p2}^l}{I_m}>L_{21} \label{fuzzycondition4}
\end{equation}
where $k'_{p1},k'_{d1},k'_{p2},k'_{d2}$ are the static parameters for cascaded PD controller without FLRs, and $min()$ means taking the minimal value. 

\textbf{Observation 4}: The stability conditions for FLR-enhanced PD control in (\ref{fuzzycondition1})-(\ref{fuzzycondition4}) are a kind of Membership-Function-Independent (MFI) method \cite{669023,481841}, by using Membership Function Boundary (MFB) techniques \cite{4565671,iet:/content/journals/10.1049/iet-cta.2010.0619}. Although those conditions are nearly "free" to be derived, they come with a great extent of conservativeness \cite{Lam2016PolynomialFM}. By considering the internal dynamics of the fuzzy logic systems, the stability conditions can be relaxed by introducing slack matrices.

\section{Simulation} \label{simulation}
This section introduces the implementation and results in simulation. Firstly, some necessary parameters for simulation are specified. Secondly, BO tuning process is detailed, which renders the parameters of the controllers. Next, the numerical results as well as stability analysis are carried out. Further, we implemented ablation experiments to illustrate the contribution of each component of our controller.

\subsection{Parameters Specification}
Tab.\ref{table_dynamic} renders the parameters for dynamic model. The values of those parameters are taken from a real physical machine \cite{BANSAL2021375}. Tab.\ref{table_param_simulation} is the basic setting for simulation environment. The initial values are $\{ x_i=0|i=1,2,3,4 \}$. Two reference signals are implemented. The first one is square-wave signal, which is defined as
\begin{equation}
x_{1d}(t)=
\begin{cases}
1, \text{if t < 10 sec} \\
0, \text{otherwise} \\
\end{cases} 
,\dot{x}_{1d}(t)=0. \label{squarewave}
\end{equation}
The other one is sine-wave target $[x_{1d}(t),\dot{x}_{1d}(t)]^T=[sin(t),cos(t)]^T$. The total disturbance $d_1,d_2$ are set as random values ranging between $[-10,10] rad/s^2$.

\begin{table}[!htbp]
\centering
\caption{Parameter of Dynamic Model}
\begin{tabular}{ccc}
\hline
Parameters & Description & Values\\
\hline
$g$ & Gravity acceleration & $9.8 m/sec^2$\\
$m$ & Mass of the link & $1.2756 kg$\\ 
$l$ & Length of the link & $0.4m$ \\
$I_l$ & Inertia of link & $1 kg \ m^2$\\
$I_m$ & Inertia of motor & $0.3 kg \ m^2$ \\
$k$ & Elastic stiffness of the flexible link & $100 Nm$ \\
$\mu$ & Viscosity & $0.1 kg \ m^2/sec$ \\
\hline \label{table_dynamic}
\end{tabular}
\end{table}

\begin{table}[!htbp]
\centering
\caption{Parameter of Simulation Environment}
\begin{tabular}{cc}
\hline
Description & Values\\
\hline
Simulation timestep & $0.005sec$\\
ODE solver & Forward Euler\\ 
Control timestep & $0.05sec$ \\
Episode & $10sec$ \\
Env & OpenAI Gym \cite{gym}\\
\hline \label{table_param_simulation}
\end{tabular}
\end{table}

\subsection{BO Tuning Procedure and Results} \label{BOTuningProcedureandResults}
BO is implemented in this paper to achieve data-efficient tuning. The advantages of BO is that it can find a sub-optimal solution quickly. In this paper, BO is first utilized to tune the parameters of two PD controllers jointly without introducing fuzzy logic regulator. The cost function is negative sum of absolute angular error, and square-wave signal (\ref{squarewave}) is used as reference. Therefore, the task of BO can be formalized:
\begin{equation}
    \max_{k_{p1},k_{d1},k_{p2},k_{d2}} -\sum_{timestep=0}^{200}|e_1| \label{cost} 
\end{equation}
The searching ranges are limited to be
\begin{equation}
    k_{p1},k_{p2} \in [0,150],k_{d1},k_{d2} \in [0,30].
\end{equation}

\begin{figure}[htbp]
\centering
\begin{minipage}[t]{0.48\textwidth}
\centering
\includegraphics[scale=0.48]{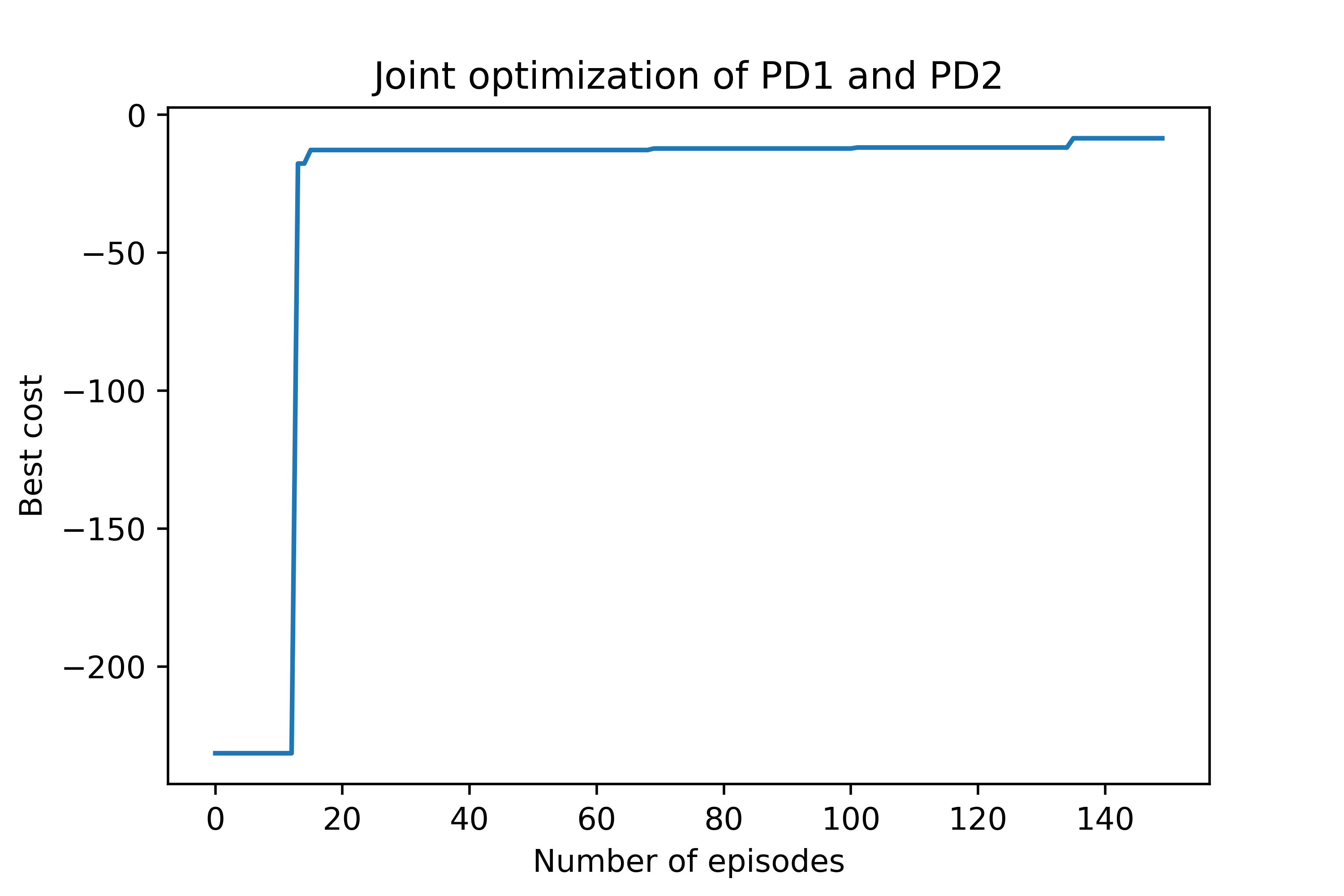}
\caption{Joint tuning of PD1 and PD2} \label{jointPDBO}
\end{minipage}
\begin{minipage}[t]{0.48\textwidth}
\centering
\includegraphics[scale=0.48]{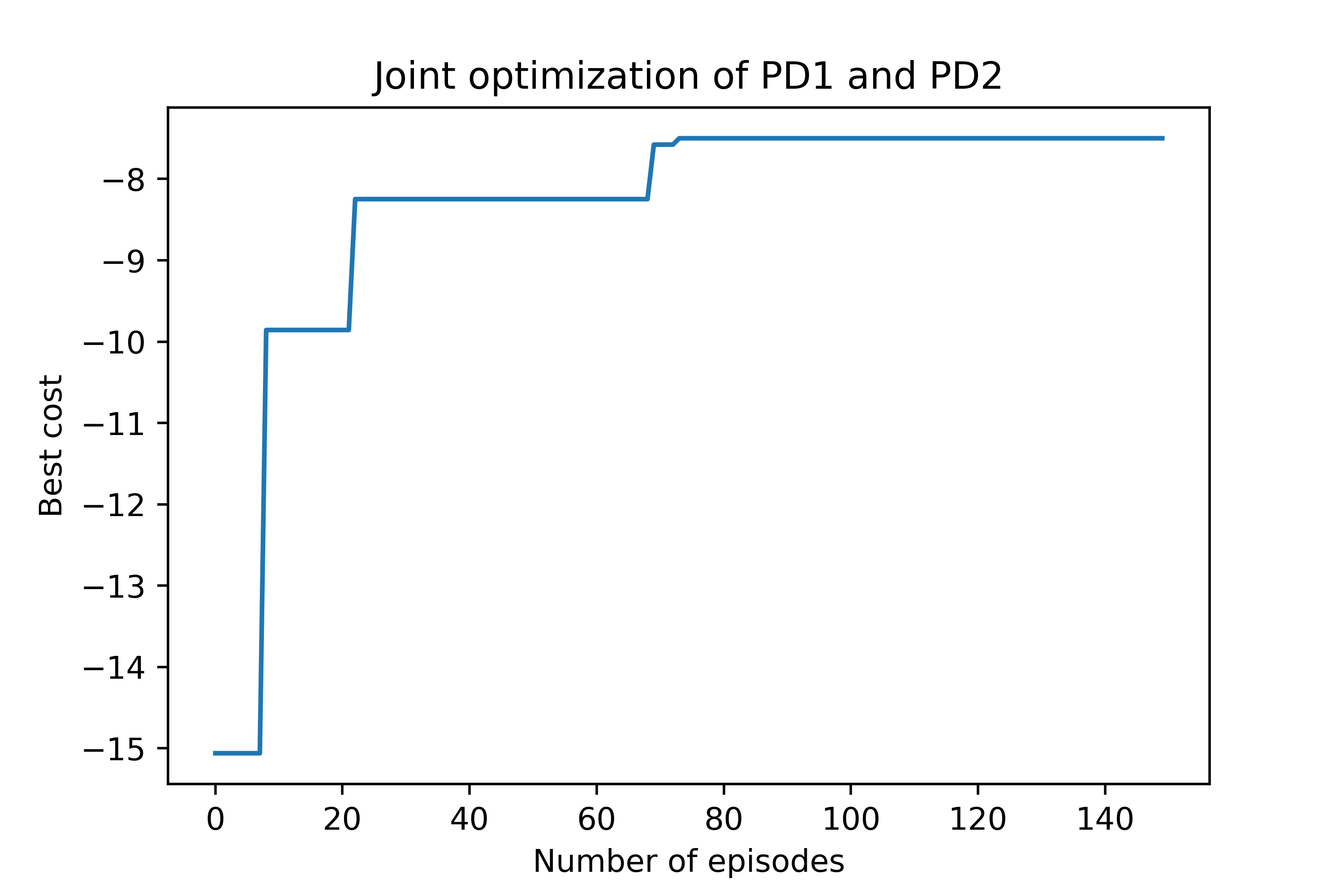}
\caption{Tuning of FLR} \label{PDfuzzyBO}
\end{minipage}
\end{figure}

BO is run for 150 episodes, and Fig.\ref{jointPDBO} records the highest cost encountered upon each number of episodes. Upon retrieving the "best" set of parameters for PD controllers, we use that set of parameters as baseline and tune the upper/lower bounds for fuzzy logic regulators. The FLR is responsible for adjusting 4 parameters, with each parameter having one upper bound and one lower bound. Therefore, BO tuning for FLR has eight parameters. Fig.\ref{PDfuzzyBO} records the highest cost encountered upon each number of episodes for FLR tuning. At last, the resulting parameters are summarized in Tab.\ref{table_BOresult}. It should be noticed that BO in practice can quickly converge to satisfying performance within a few iterations. This is valuable to practice, since it means a satisfying set of parameters is easily accessible with low damage to the devices.

\begin{table}[!htbp]
\centering
\caption{Tuning Result of BO}
\begin{tabular}{ccc}
\hline
Parameters & Description & Values\\
\hline
$k_{p1}$ & Proportional gain for PD1 & $52.19$ \vspace{1ex}\\
$k_{d1}$ & Derivative gain for PD1 & $10.18$ \vspace{1ex}\\ 
$k_{p2}$ & Proportional gain for PD2 & $144.5$\vspace{1ex} \\
$k_{d2}$ & Derivative gain for PD2 & $8.636$ \vspace{1ex}\\
$\Delta k_{p1}^u$ & Upper bound of FLR on $k_{p1}$ & $-11.61$ \vspace{1ex} \\
$\Delta k_{p1}^l$ & Lower bound of FLR on $k_{p1}$ & $15.27$ \vspace{1ex} \\
$\Delta k_{d1}^u$ & Upper bound of FLR on $k_{d1}$ & $-3.228$ \vspace{1ex} \\
$\Delta k_{d1}^l$ & Lower bound of FLR on $k_{d1}$ & $0.1$ \vspace{1ex} \\
$\Delta k_{p2}^u$ & Upper bound of FLR on $k_{p2}$ & $-16.94$ \vspace{1ex} \\
$\Delta k_{p2}^l$ & Lower bound of FLR on $k_{p2}$ & $2.997$ \vspace{1ex} \\
$\Delta k_{d2}^u$ & Upper bound of FLR on $k_{d2}$ & $-0.1$ \vspace{1ex} \\
$\Delta k_{d2}^l$ & Lower bound of FLR on $k_{d2}$ & $0.9537$ \vspace{1ex} \\
\hline \label{table_BOresult}
\end{tabular}
\end{table}

\subsection{Results and Evaluation}

\subsubsection{Stability Analysis of Cascaded PD Controller} \label{StabilityAnalysisofCascadedPDController}
\ 

With the parameters given in Tab.\ref{table_BOresult} and Tab.\ref{table_dynamic}, the stability analysis can be carried out both in terms of transfer function and Jacobian matrix. In this paper, only cascade PD controller without fuzzy logic regulators is analyzed, since the stability analysis of fuzzy logic system itself is still a heated point of research. The fuzzy cascaded PD controller will be investigated numerically in the following sections.

Substituting all the parameters into (\ref{tf11d}, we have the final transfer function of $x_1$ w.r.t $x_{1d}$
\begin{equation}
    \frac{X_1}{X_{1d}}= \frac{531.2942s+12760.455}{0.3s^4+18.636s^3+274.5s^2+1624.5846s+12676.455},\label{tf11d_}
\end{equation}  
of which the poles are solved as 
\begin{equation}
    p_1=-43.3921,
\end{equation}
\begin{equation}
    p_2=-16.1253,
\end{equation}
\begin{equation}
    p_3=-1.3013 + 7.6613i,
\end{equation}
\begin{equation}
    p_4=-1.3013 - 7.6613i.
\end{equation}

Evidently, because all poles are to the left of imaginary axis, the system is stable. Further, two of the poles are in the real axis.

As for the Jacobian matrix analysis, substituting all the parameters into (\ref{DF}). Note that the reference signal is only dependent on time $t$, so $\frac{\partial \ddot{x}_{1d}}{\partial e_i}=0,i=1,2,3,4$. Similarly, the disturbances are random values in this paper, therefore $\frac{\partial d_i}{\partial e_i}=0,i=1,2,j=1,2,3,4$. Finally, the Jacobian matrix becomes
\begin{equation}
DF(e_1,e_2,e_3,e_4)=
\left [
\begin{array}{cccc}
    0 & 1 & 0 & 0 \\
    -52.19 & 
    -10 & 
    100 & 
    0 \\
    0 & 0 & 0 & 1 \\
    0 & 
    0 & 
    -815 & 
    -29.12
\end{array}
\right ] ,
\end{equation}
and the eigenvalues are 
\begin{equation}
    \lambda_1=-5.0000 + 5.2144i,
\end{equation}
\begin{equation}
     \lambda_2=-5.0000 - 5.2144i,
\end{equation}
\begin{equation}
     \lambda_3=-14.5600 +24.5562i,
\end{equation}
\begin{equation}
     \lambda_4=-14.5600 -24.5562i.
\end{equation}

Similarly, all the eigenvalue of $DF$ have negative real part, which ensures our system is asymptotically stable. After introducing FLRs, in the worst-case scenario, the Jacobian matrix becomes 
\begin{equation}
DF(e_1,e_2,e_3,e_4)=
\left [
\begin{array}{cccc}
    0 & 1 & 0 & 0 \\
    -40.58 & 
    -6.772 & 
    100 & 
    0 \\
    0 & 0 & 0 & 1 \\
    0 & 
    0 & 
    -758.53 & 
    -28.79
\end{array}
\right ],
\end{equation}
of which the eigenvalues are 
\begin{equation}
    \lambda_1=-3.3860 + 5.3958i,
\end{equation}
\begin{equation}
     \lambda_2=-3.3860 - 5.3958i,
\end{equation}
\begin{equation}
     \lambda_3=-14.3950 +23.4801i,
\end{equation}
\begin{equation}
     \lambda_4=-14.3950 -23.4801i.
\end{equation}
of which all eigenvalues have negative real parts. This illustrates the stability conditions (\ref{fuzzycondition1})-(\ref{fuzzycondition4}) are satisfied.

\subsubsection{Square-wave Signal Tracking} 
\ 

\begin{figure}[htbp]
\centering
\begin{minipage}[t]{0.48\textwidth}
\centering
\includegraphics[scale=0.48]{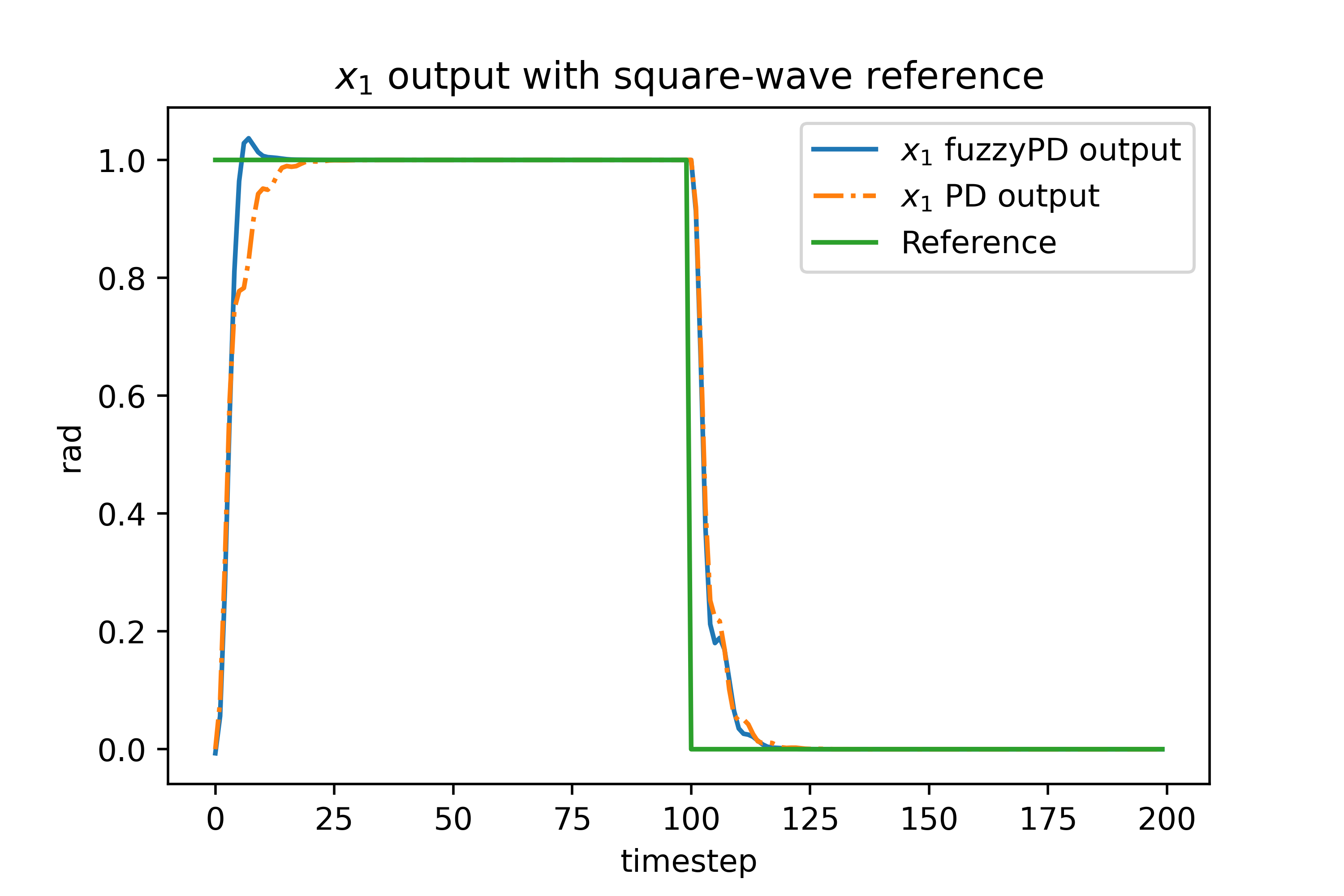}
\caption{$x_1$ output with square-wave reference} \label{x1square}
\end{minipage}
\begin{minipage}[t]{0.48\textwidth}
\centering
\includegraphics[scale=0.48]{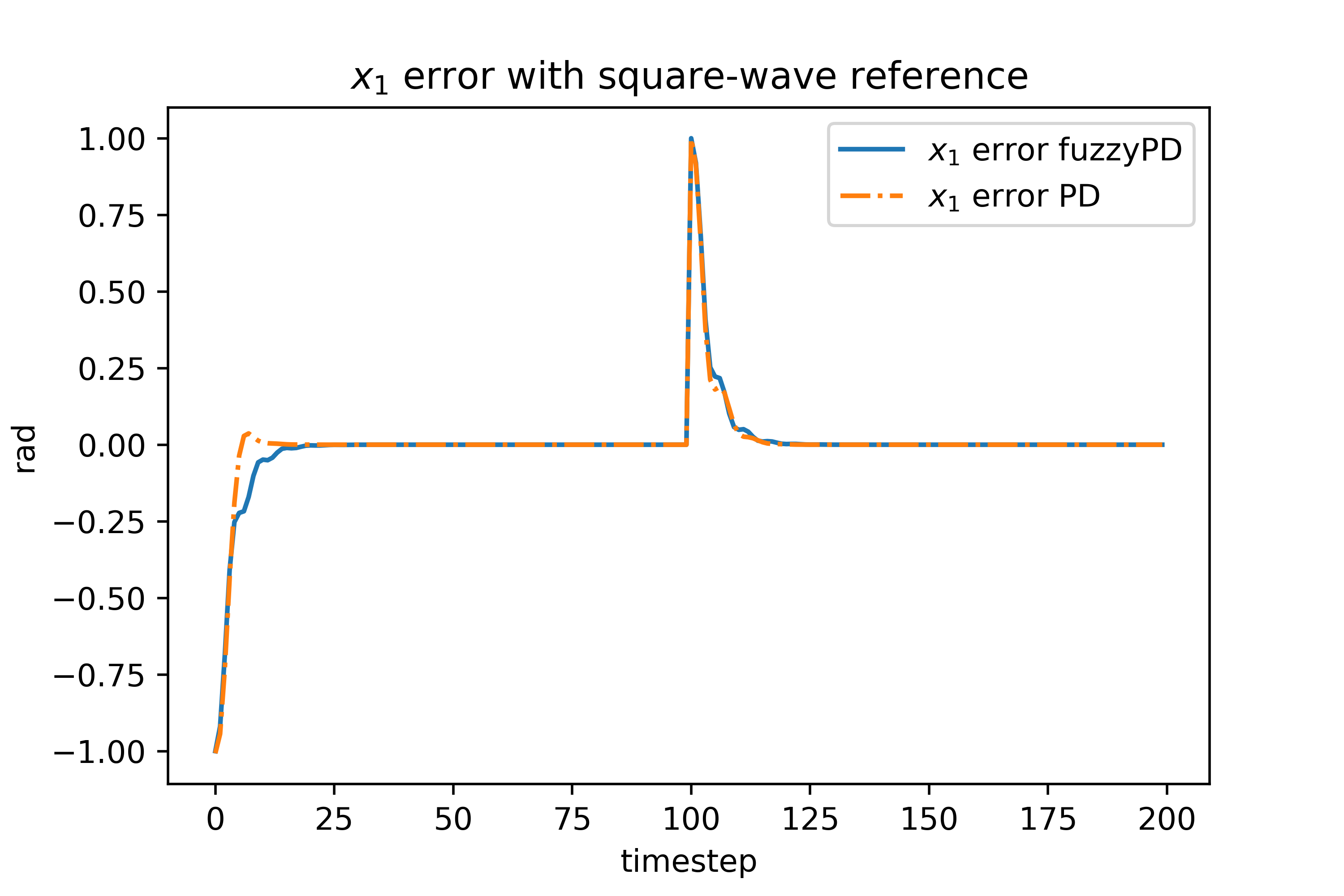}
\caption{$x_1$ error with square-wave reference} \label{x1errorsquare}
\end{minipage}
\end{figure}

\begin{figure}[htbp]
\centering
\begin{minipage}[t]{0.48\textwidth}
\centering
\includegraphics[scale=0.48]{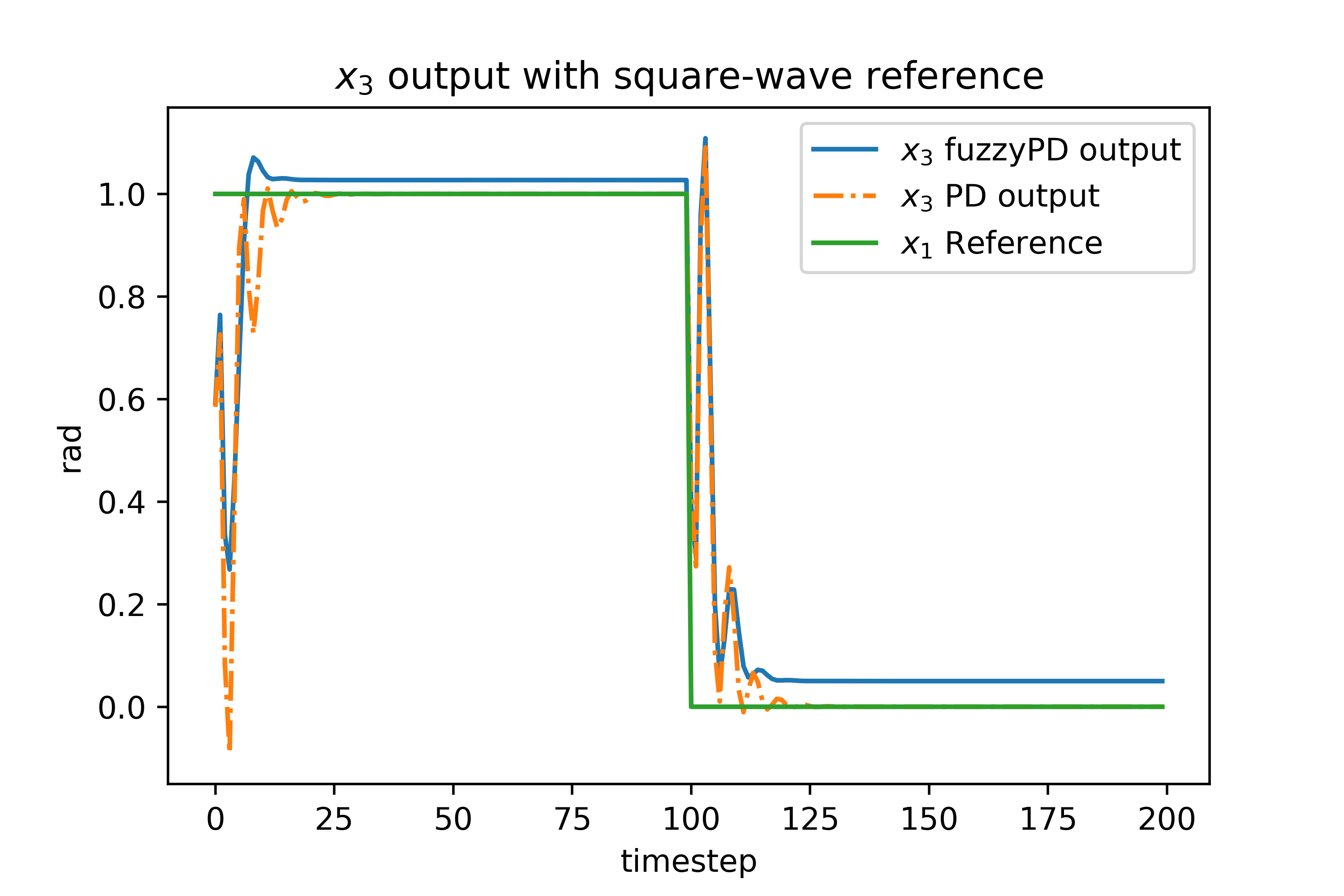}
\caption{$x_3$ output with square-wave reference} \label{x3square}
\end{minipage}
\begin{minipage}[t]{0.48\textwidth}
\centering
\includegraphics[scale=0.48]{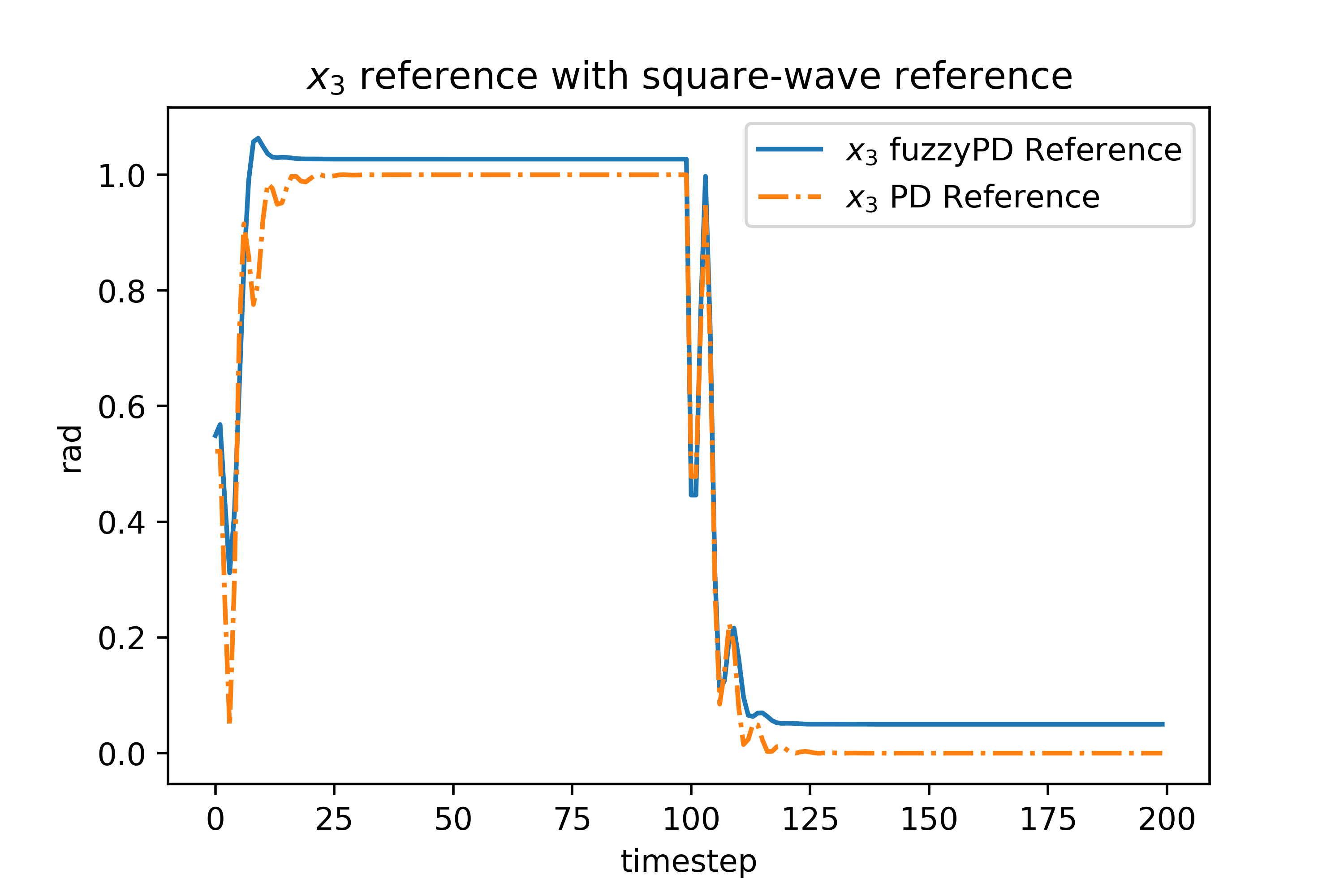}
\caption{$x_3$ reference with square-wave reference} \label{x3refsquare}
\end{minipage}
\end{figure}

\begin{figure}[htbp]
\centering
\begin{minipage}[t]{0.48\textwidth}
\centering
\includegraphics[scale=0.48]{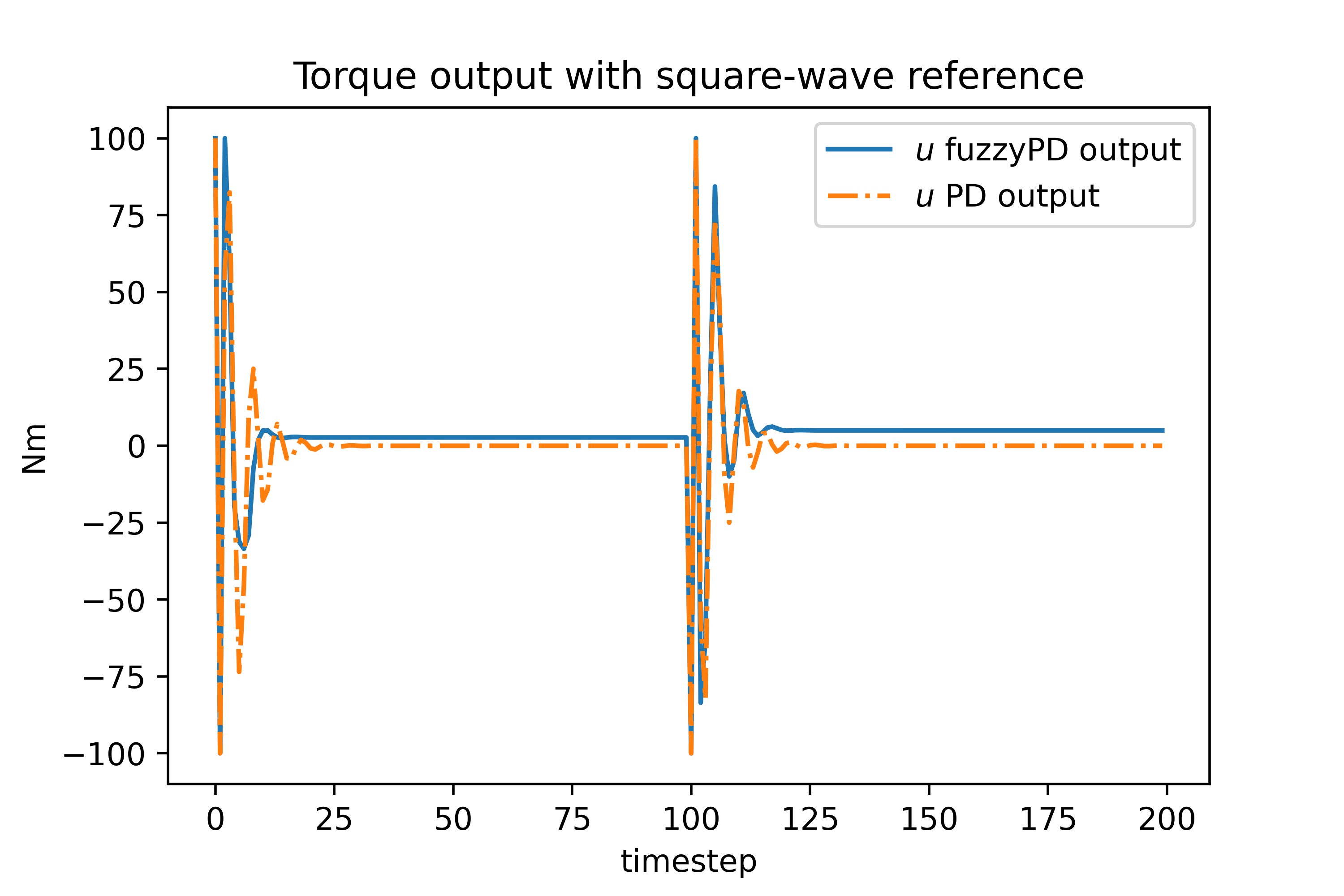}
\caption{Torque input with square-wave reference} \label{usquare}
\end{minipage}
\begin{minipage}[t]{0.48\textwidth}
\centering
\includegraphics[scale=0.48]{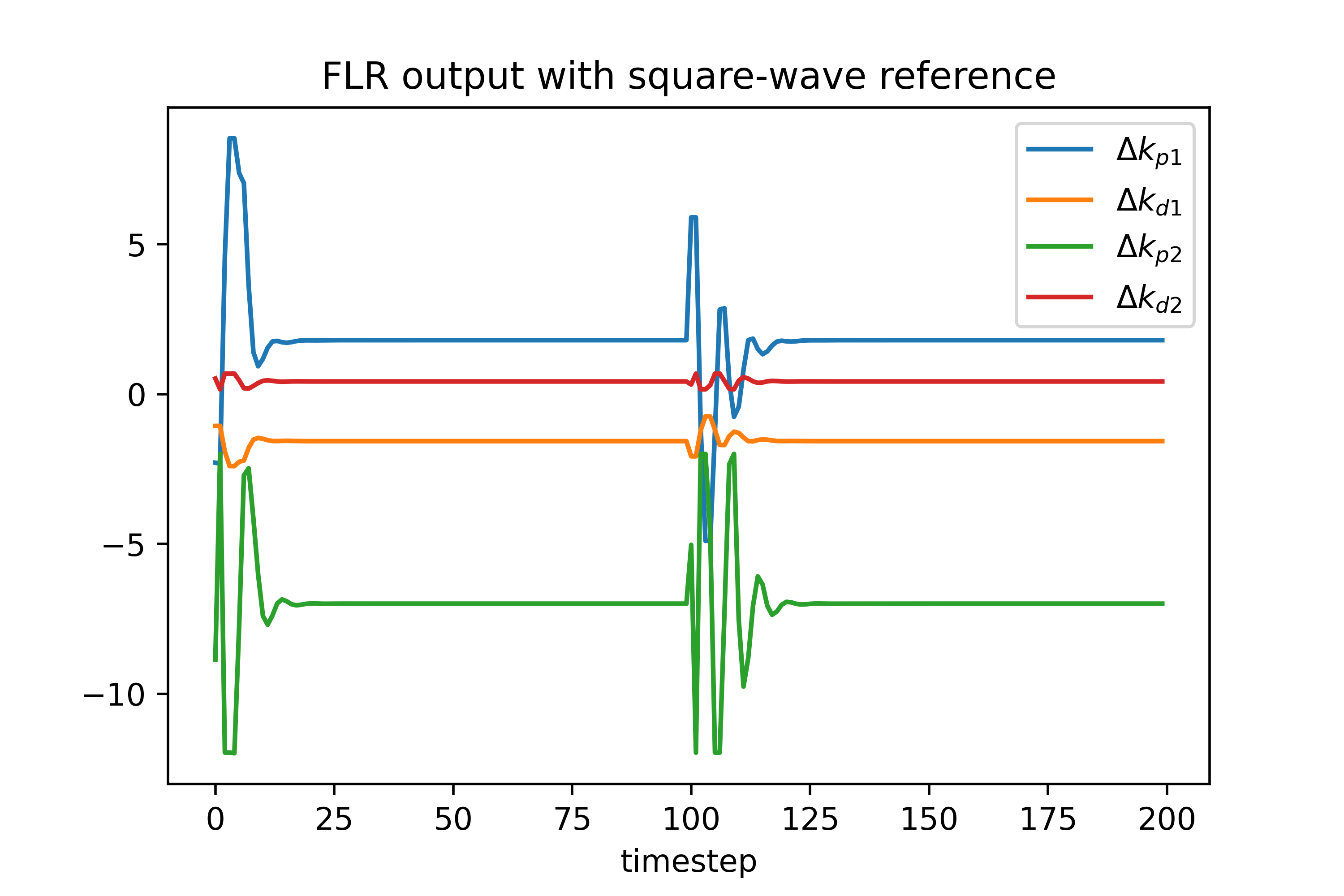}
\caption{FLR output with square-wave reference} \label{fuzzyoutsquare}
\end{minipage}
\end{figure}

The main results of square-wave signal tracking are presented in Fig.\ref{x1square} to Fig.\ref{fuzzyoutsquare}. Fig.\ref{x1square} and Fig.\ref{x1errorsquare} are the $x_1$ output and error profile respectively. In the legend, "fuzzyPD" means fuzzy cascaded PD controller proposed in this paper, and "PD" represents conventional cascaded PD without fuzzy logic regulators. We can see that the main difference is that "fuzzyPD" has shorter settling time, but with the cost of $3.68\%$ overshoot. Besides, it is noticeable that conventional cascaded PD controller here can already achieve smooth motion without overshoot. Fig.\ref{x3square} and Fig.\ref{x3refsquare} are the $x_3$ outputs and references. Both controllers shows oscillation during the reaching phase, but the trajectory of "fuzzyPD" is smoother in comparison. Besides, near the equilibrium, the $x_3$ reference of "fuzzyPD" is larger than "PD", which greatly reduces the equilibrium error from $3.9 \times 10^{-5}$ to $-7.15 \times 10^{-7}$. Fig.\ref{usquare} and Fig.\ref{fuzzyoutsquare} represent the torque and FLR outputs. The profile of the torque follows the same trend of $x_3$ outputs. Similar oscillations are witnessed during the reaching phase, and higher torque with "fuzzyPD" near the equilibrium. In Fig.\ref{fuzzyoutsquare}, one interesting fact is that $\Delta K_{p}$ and $\Delta K_{d}$ have opposite rate of change. When $\Delta K_{p}$ is increasing, $\Delta K_{d}$ is decreasing. This coincides with the design process of fuzzy logic regulator. Also, FLR for PD1 tries to increase the response speed by increasing $\Delta K_{p1}$ and decreasing $\Delta K_{d1}$, while it is just the opposite for PD2. 

\subsubsection{Sine-wave Signal Tracking}
\ 

\begin{figure}[htbp]
\centering
\begin{minipage}[t]{0.48\textwidth}
\centering
\includegraphics[scale=0.48]{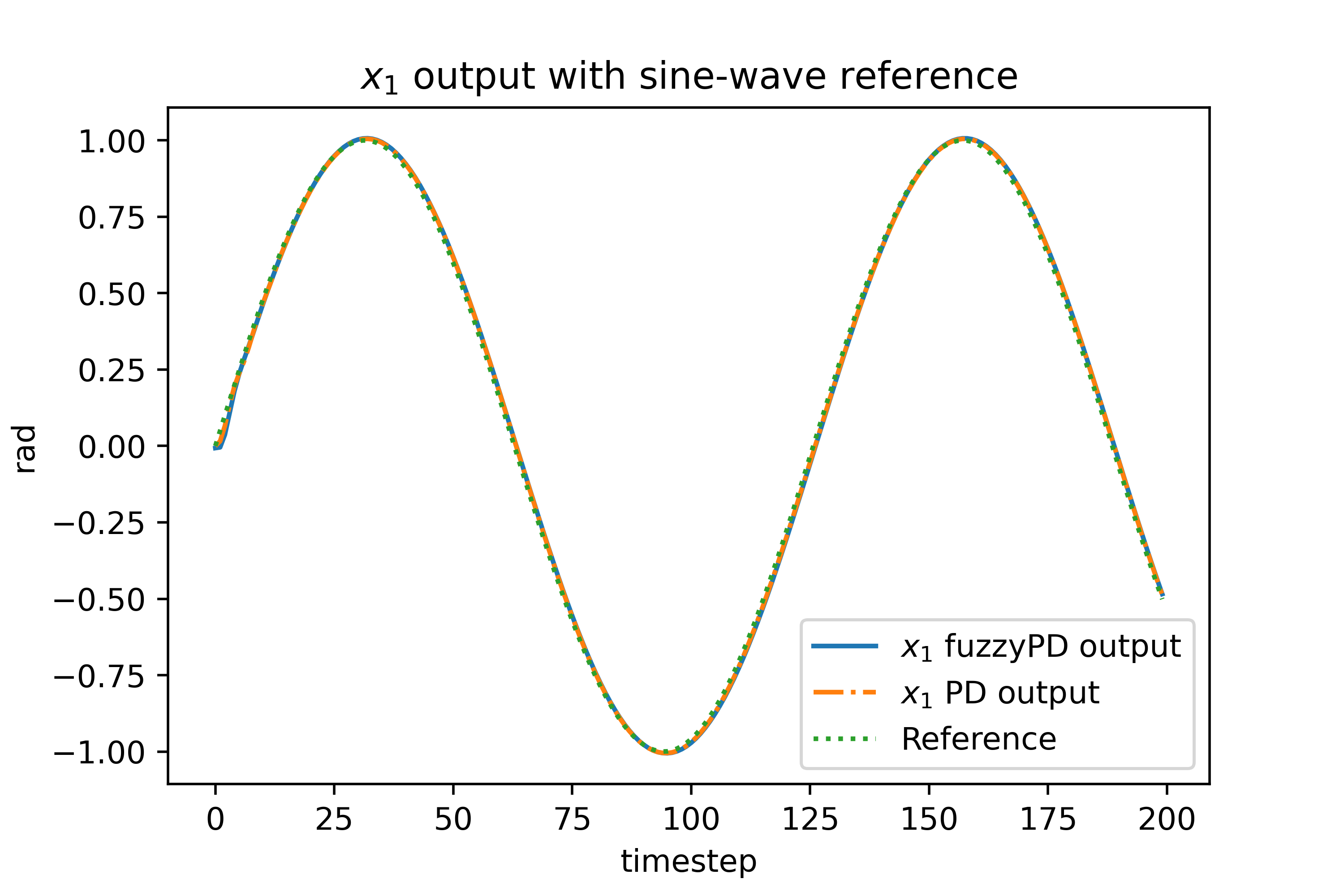}
\caption{$x_1$ output with sine-wave reference} \label{x1sine}
\end{minipage}
\begin{minipage}[t]{0.48\textwidth}
\centering
\includegraphics[scale=0.48]{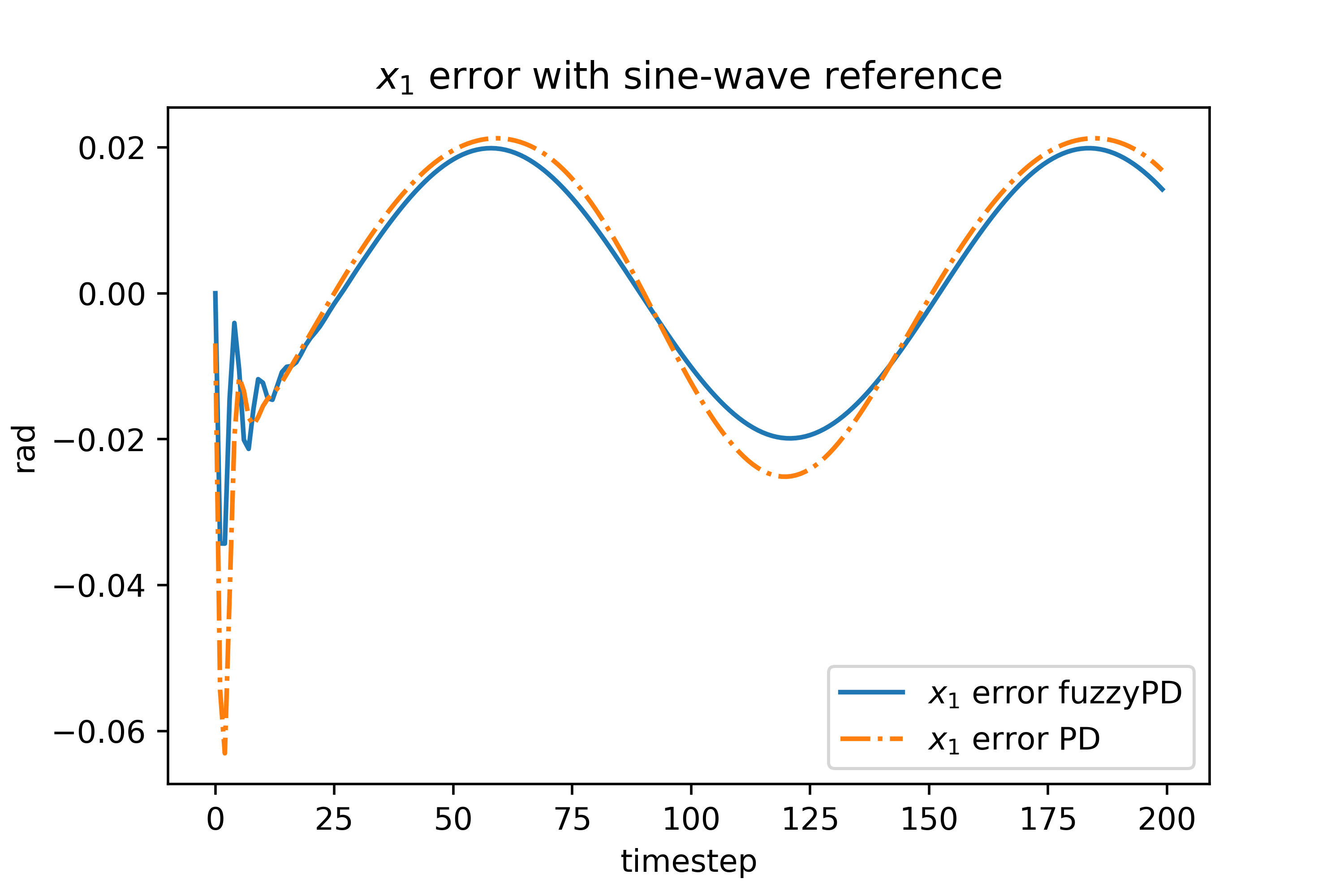}
\caption{$x_1$ error with sine-wave reference} \label{x1errorsine}
\end{minipage}
\end{figure}

\begin{figure}[htbp]
\centering
\begin{minipage}[t]{0.48\textwidth}
\centering
\includegraphics[scale=0.48]{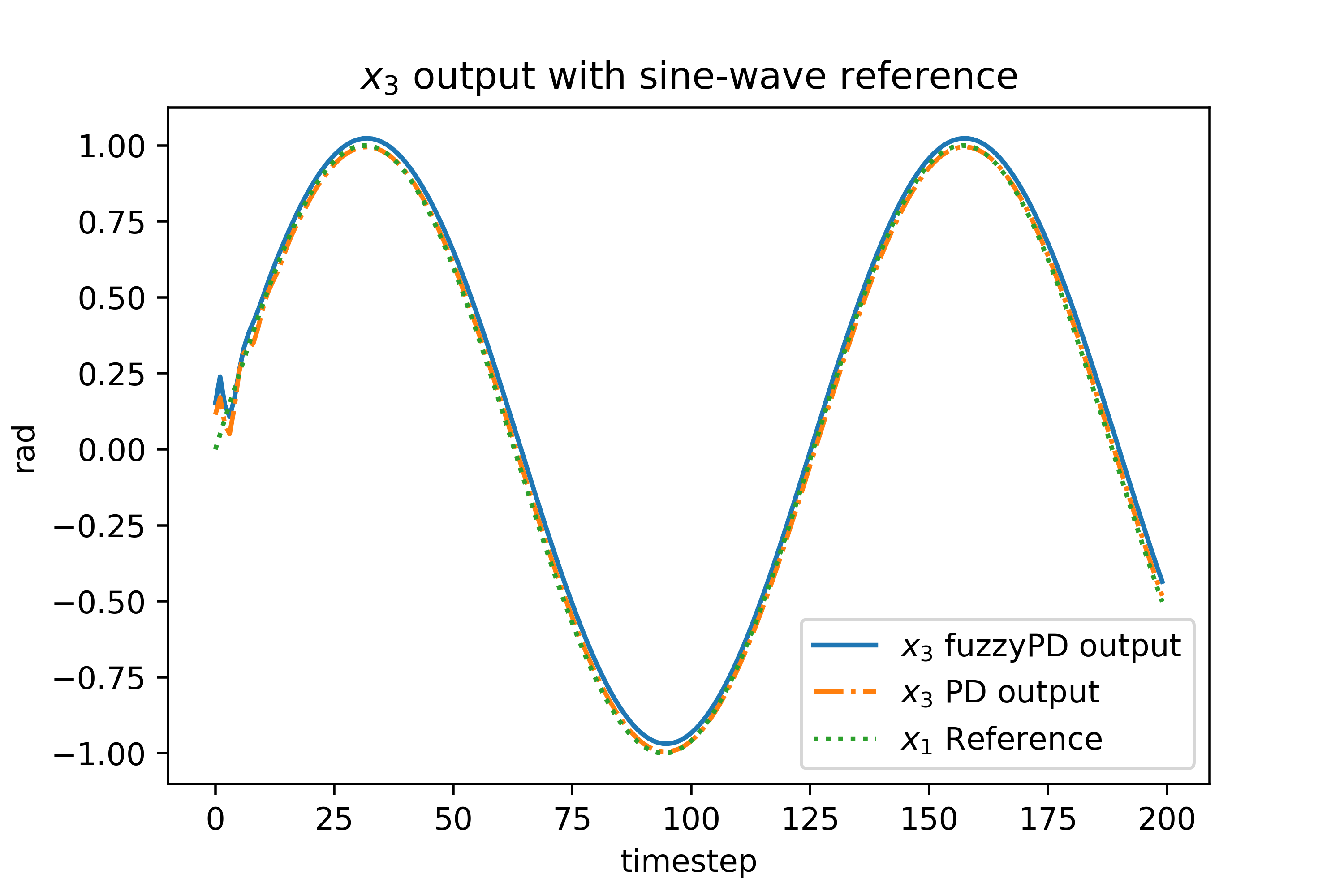}
\caption{$x_3$ output with sine-wave reference} \label{x3sine}
\end{minipage}
\begin{minipage}[t]{0.48\textwidth}
\centering
\includegraphics[scale=0.48]{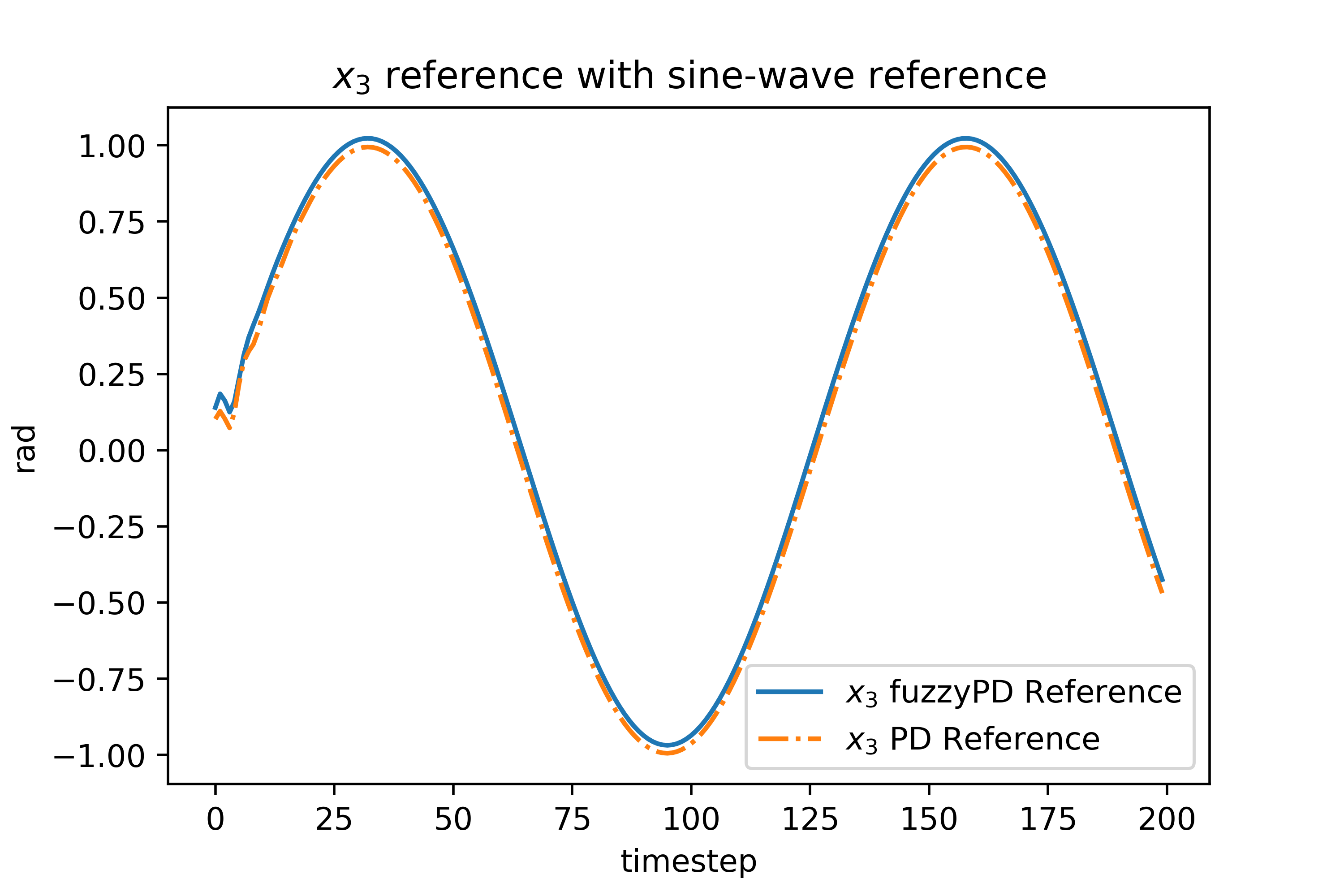}
\caption{$x_3$ reference with sine-wave reference} \label{x3refsine}
\end{minipage}
\end{figure}

\begin{figure}[htbp]
\centering
\begin{minipage}[t]{0.48\textwidth}
\centering
\includegraphics[scale=0.48]{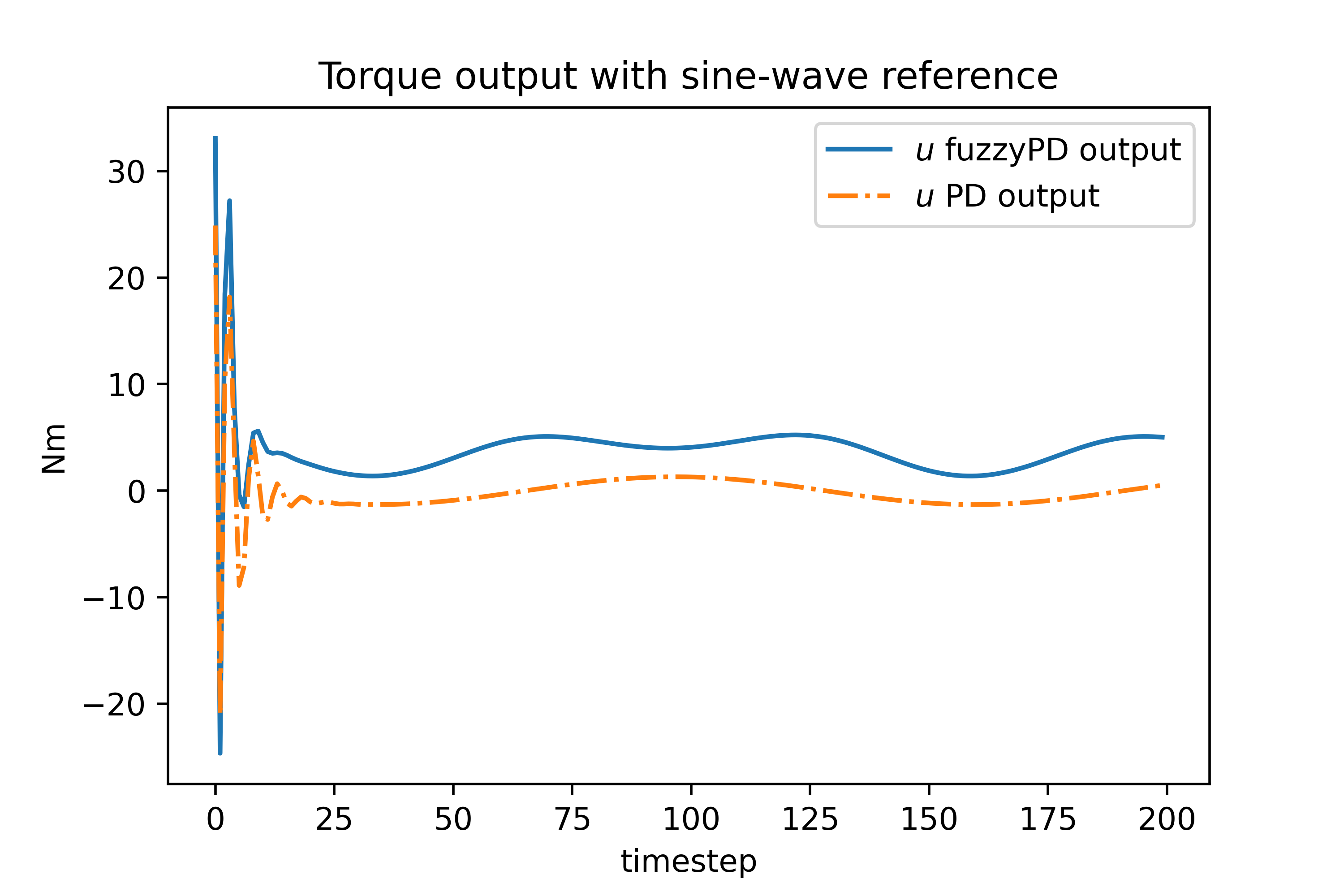}
\caption{Torque input with sine-wave reference} \label{usine}
\end{minipage}
\begin{minipage}[t]{0.48\textwidth}
\centering
\includegraphics[scale=0.48]{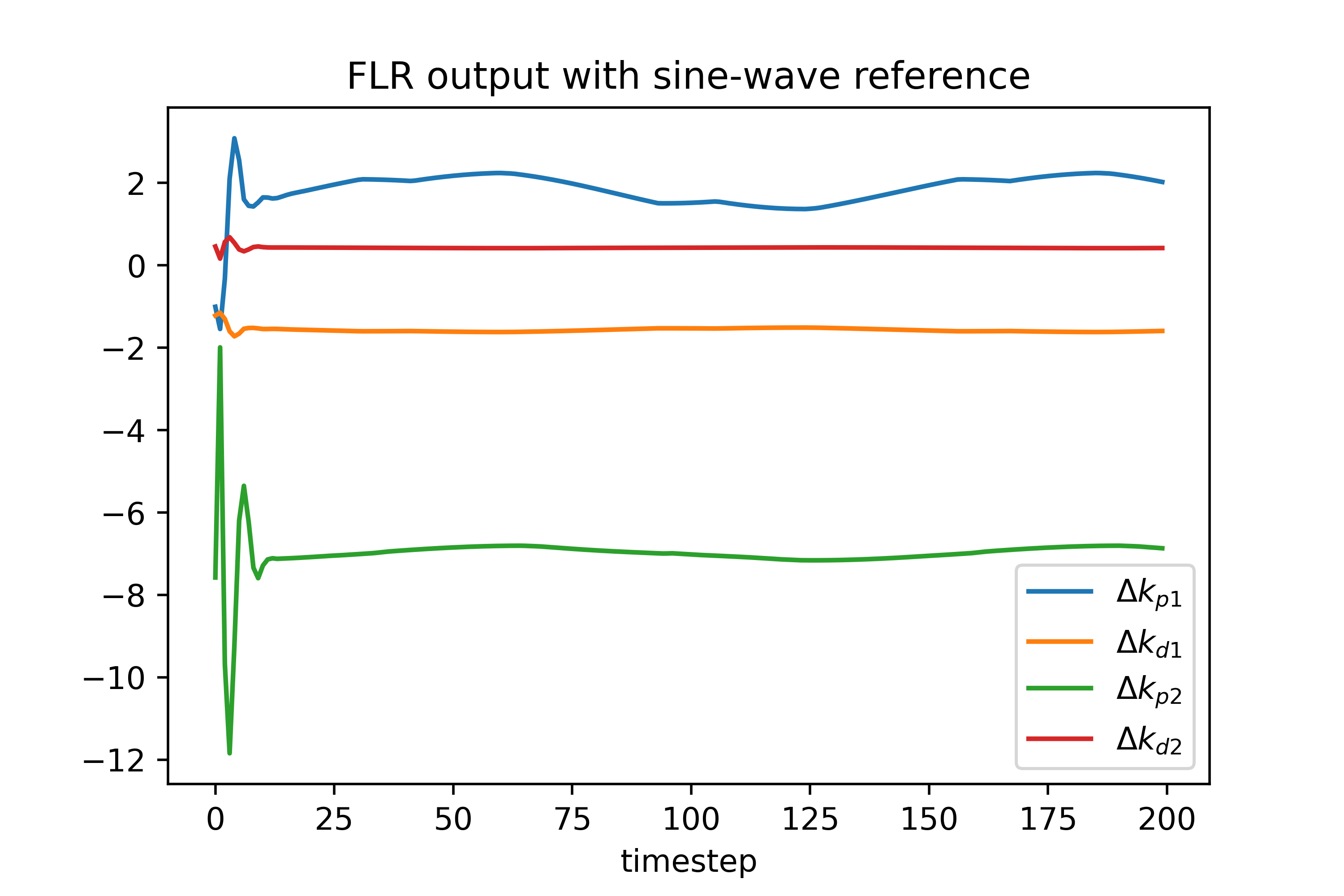}
\caption{FLR output with sine-wave reference} \label{fuzzyoutsine}
\end{minipage}
\end{figure}

The main results of sine-wave signal tracking are presented in Fig.\ref{x1sine} to Fig.\ref{fuzzyoutsine}. Fig.\ref{x1sine} and Fig.\ref{x1errorsine} are the $x_1$ output and error profile respectively. The tracking performance is satisfying for both controllers, but the error profile illustrates that "fuzzyPD" has overall smaller error. Fig.\ref{x3sine} and Fig.\ref{x3refsine} are the $x_3$ outputs and references. Conceivably, $x_3$ tracks its desired path relatively well.  Fig.\ref{usine} and Fig.\ref{fuzzyoutsine} represent the torque and FLR outputs. While the torque of "PD" follows the trend of sine wave, the torque of "fuzzyPD" shows more complicated pattern. We believe this alternation helps "fuzzyPD" to maintain low errors. For FLR outputs, because the tracking errors are not changing rapidly, the outputs of FLR are almost constant. Besides, it has similar behavior with that in square-wave tracking task.

\subsection{Ablation experiment}
In this section, some ablation experiments are implemented to investigate how each component of our proposed controller contribute to the final results. For simplicity, only $x_1$ output with square-wave reference is illustrated.

\begin{figure}[htbp]
\centering
\begin{minipage}[t]{0.48\textwidth}
\centering
\includegraphics[scale=0.5]{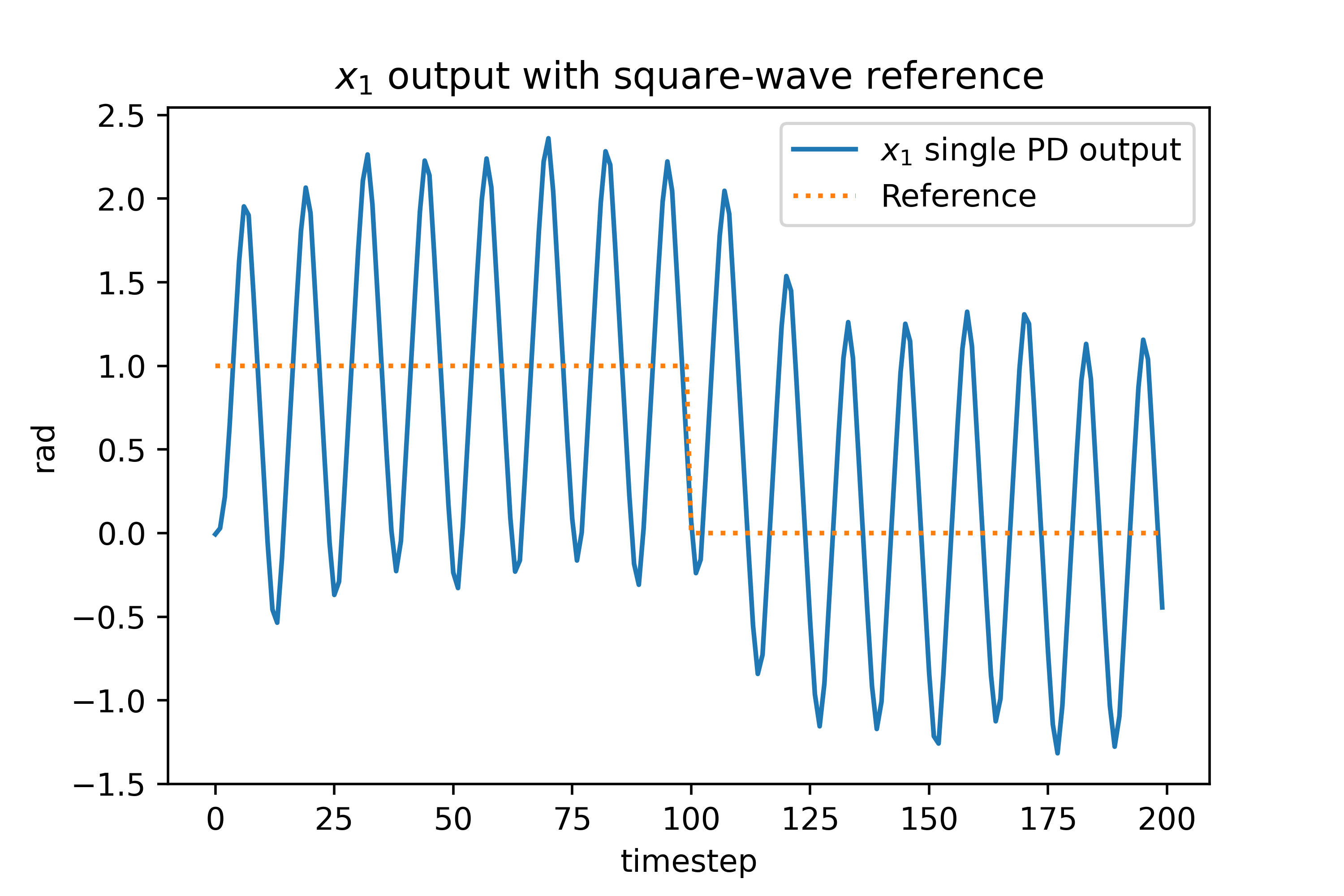}
  \caption{Output of single PD controller}
  \label{singlePD}
\end{minipage}
\begin{minipage}[t]{0.48\textwidth}
\centering
\includegraphics[scale=0.5]{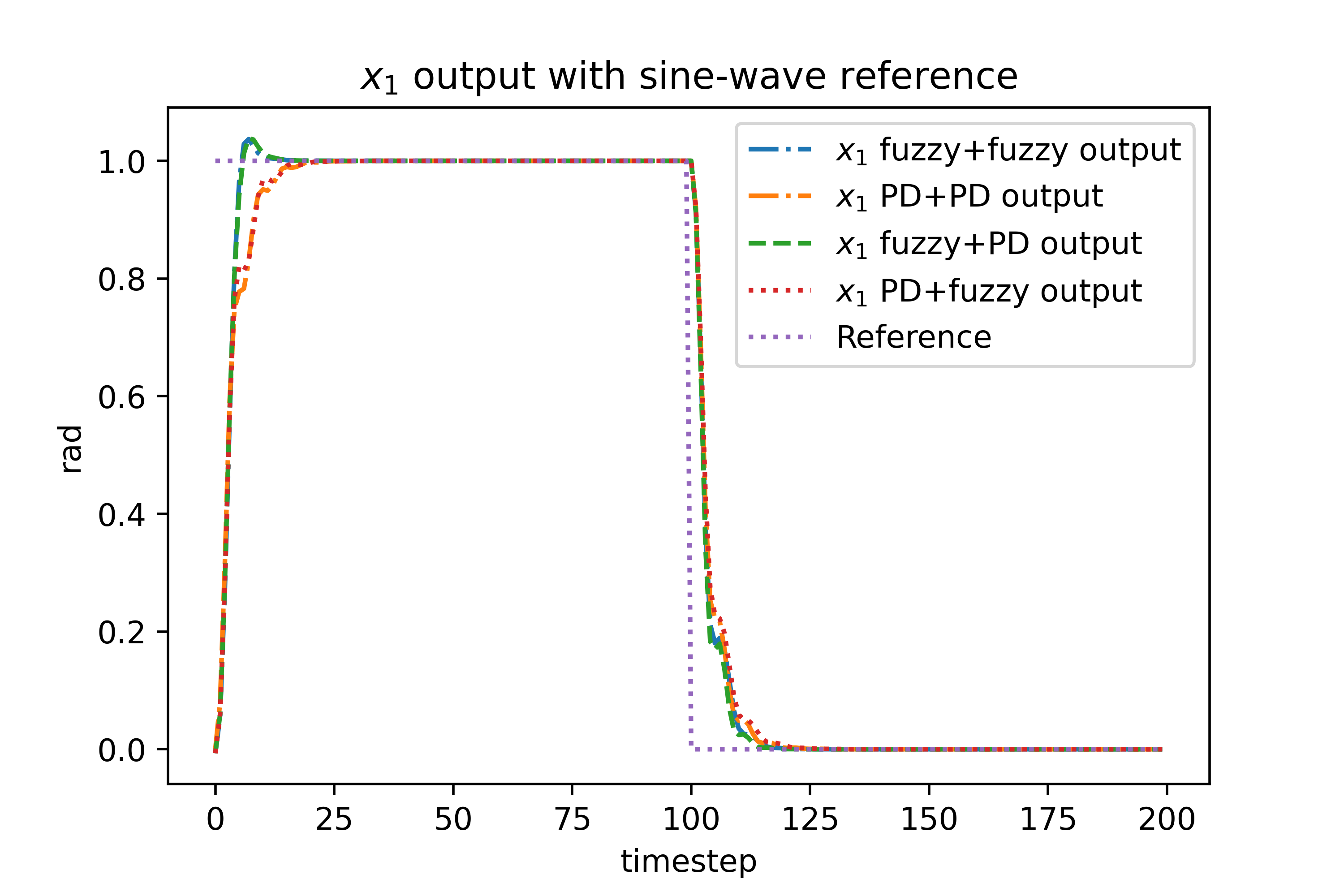}
  \caption{Outputs of ablation study}
  \label{ablation}
\end{minipage}
\end{figure}

Firstly, as a baseline, one single PD controller is tuned by BO. The resulting parameters are $[k_p,k_d]=[117.0,29.99]$. The $x_1$ output is shown in Fig.\ref{singlePD}. Obviously, single PD controller behaves poorly here, with nearly constant magnitude oscillation. It is understandable, since single PD controller here is in essence just a reduced-order controller.

\begin{table}[!htbp]
\centering
\caption{Cost for ablation experiments}
\begin{tabular}{ccccc}
\hline
Controllers & fuzzy+fuzzy & PD+PD & fuzzy+PD & PD+fuzzy\\
\hline
Cost/rad & -7.50 & -8.54 & -7.25 & -8.71\\

\hline \label{ablation_tab}
\end{tabular}
\end{table}

Further, we implemented 4 ablation experiments, of which the outputs are depicted together in Fig.\ref{ablation}, and the cost values are summarized in Tab.\ref{ablation_tab}. The cost is calculated following (\ref{cost}). "fuzzy+fuzzy" means two fuzzy PD controllers are used; "PD+PD" means two conventional PD controllers; "fuzzy+PD" represents fuzzy PD for sub-plant1 and conventional PD for sub-plant2; "PD+fuzzy" represents conventional PD for sub-plant1 and fuzzy PD for sub-plant2. It is evident that "fuzzy+fuzzy" and "fuzzy+PD" have similar performance, and are better than the other two. The rationale behind is that the performance of the first PD controller is more critical than the second. The first PD controller is used to specify a smooth yet rapid convergence path to the reference. while the second PD controller tracks that smooth path until convergence. For a PD controller tracking a smooth path is an easy task, which explains why a fuzzy PD2 controller is not improving the performance much. Besides, we witness that "fuzzy+PD" has slightly better performance than "fuzzy+fuzzy". We believe it is caused by BO getting stuck in a local minimum, or that the number of optimization episodes is not large enough. In addition, we should note that lower cost in this paper does not really mean "better". It only means that the system converges quickly, but maybe at the cost of overshoot. From Fig.\ref{fuzzyoutsquare}, we can see that $k_{p2}$ is constantly being decreased, which means FLR tries to lower the overshoot with the price of lower convergence rate, and therefore higher cost value. Further improvements may be possible by using type-2 fuzzy logic system, which is generally more powerful than its type-1 counterpart.

\section{Conclusion} \label{conclusion}
A fuzzy cascaded PD controller applied to flexible joint manipulators is proposed in this paper. The flexible joint manipulator system is a $4^{th}$-order under-actuated dynamic system, where we tried experimentally to stabilize it using single PD controller in vain. Therefore, in this paper, the $4^{th}$-order system is considered as two coupled $2^{nd}$-order sub-systems, and two PD controllers are used to control each of them separately. To derive the cascaded PD controller, the coupling terms in the first sub-system serves as a design variable. After combining with the first PD controller, the coupling term is transformed to reference signal for the second sub-system, where the second PD controller can then be implemented directly. In this case, the cascaded PD controller maintains the simplicity and explainability of conventional PID controller. Besides that, fuzzy logic systems are implemented as regulators to improve the performance of conventional PD controllers. The proposed fuzzy PD controller is proved to be asymptotically stable using Jacobian matrix and transfer function respectively. The experiments show that the cascaded PD controller fulfills the tracking task quite well with little oscillation, while the addition of fuzzy logic regulator increases the convergence speed and further cancels the oscillation. All the parameters are tuned by Bayesian Optimization, which finds satisfying results within only dozens of iterations.

For future work, there are a few points worth researching into. Firstly, to further explore the potential of fuzzy logic system, type-2 FLR could be chosen. It has been proved to be more powerful than type-1 generally. Secondly, the proposed method should be extended to MIMO systems and real robot experiments should be carried out. Last but not least, the stability condition (\ref{fuzzycondition1})-(\ref{fuzzycondition4}) are based on the lower bounds of FLR, which may be very conservative. A more relaxed stability condition should be derived by considering the dynamics of the FLR itself. For example, polynomial fuzzy logic systems with stability guarantee using LMI (Linear Matrix Inequality) and SOS (Sum of Square) \cite{Lam2016PolynomialFM} are promising and well-established methods for stable fuzzy-enhanced controller design.

\bibliographystyle{unsrt}  
\bibliography{references}  

\begin{thebibliography}{10}

\bibitem{Khoygani2015IntelligentNO}
Mohammad Reza~Rahimi Khoygani, Reza~Hasanzadeh Ghasemi, and Ahmad~Reza Vali.
\newblock Intelligent nonlinear observer design for a class of nonlinear discrete-time flexible joint robot.
\newblock {\em Intelligent Service Robotics}, 8:45--56, 2015.

\bibitem{Fateh2012NonlinearCO}
Mohammad~Mehdi Fateh.
\newblock Nonlinear control of electrical flexible-joint robots.
\newblock {\em Nonlinear Dynamics}, 67:2549--2559, 2012.

\bibitem{Rachidi2014ProportionalIntegralSM}
Mohammed Rachidi and Badr~Bououlid Idrissi.
\newblock Proportional-integral sliding mode control for trajectory tracking and vibration control of a flexible single link manipulator.
\newblock {\em International Journal of Control and Automation}, 7:203--216, 2014.

\bibitem{Yan2020TrackingCO}
Ze~Yan, Xuzhi Lai, Qingxin Meng, Qingxin Meng, P.~Zhang, P.~Zhang, and Min Wu.
\newblock Tracking control of single‐link flexible‐joint manipulator with unmodeled dynamics and dead zone.
\newblock {\em International Journal of Robust and Nonlinear Control}, 31:1270 -- 1287, 2020.

\bibitem{Yan2021ANR}
Ze~Yan, Xuzhi Lai, Qingxin Meng, and Min Wu.
\newblock A novel robust control method for motion control of uncertain single-link flexible-joint manipulator.
\newblock {\em IEEE Transactions on Systems, Man, and Cybernetics: Systems}, 51:1671--1678, 2021.

\bibitem{Dian2019AdaptiveBC}
Songyi Dian, Yi~Hu, Tao Zhao, and Jixia Han.
\newblock Adaptive backstepping control for flexible-joint manipulator using interval type-2 fuzzy neural network approximator.
\newblock {\em Nonlinear Dynamics}, 2019.

\bibitem{Yang2020ANC}
Guocai Yang, Yechao Liu, and Ming-He Jin.
\newblock A new control method of flexible-joint manipulator with harmonic drive.
\newblock {\em Proceedings of the Institution of Mechanical Engineers, Part C: Journal of Mechanical Engineering Science}, 234:1868 -- 1883, 2020.

\bibitem{Jie2021FlexibleJM}
Chen Jie.
\newblock Flexible joint manipulator controlling algorithm based on neural network improved pid.
\newblock {\em 2021 13th International Conference on Measuring Technology and Mechatronics Automation (ICMTMA)}, pages 419--422, 2021.

\bibitem{Dehghani2015FuzzyLS}
Ali Dehghani and Hamed Khodadadi.
\newblock Fuzzy logic self-tuning pid control for a single-link flexible joint robot manipulator in the presence of uncertainty.
\newblock {\em 2015 15th International Conference on Control, Automation and Systems (ICCAS)}, pages 186--191, 2015.

\bibitem{Dehghani_khodadadi_2016}
Ali Dehghani and Hamed khodadadi.
\newblock Self-tuning pid controller design using fuzzy logic for a single-link flexible joint robot manipulator.
\newblock {\em Jurnal Teknologi}, 78(6-13), Jun. 2016.

\bibitem{Yang2021AMP}
Guocai Yang, Yechao Liu, Junhong Ji, Ming-He Jin, and Songhao Piao.
\newblock A model-based pid-like motion control method for flexible-joint manipulator with harmonic drives.
\newblock {\em Proceedings of the Institution of Mechanical Engineers, Part C: Journal of Mechanical Engineering Science}, 235:7880 -- 7893, 2021.

\bibitem{NeumannBrosig2020DataEfficientAW}
Matthias Neumann-Brosig, Alonso Marco, Dieter Schwarzmann, and Sebastian Trimpe.
\newblock Data-efficient autotuning with bayesian optimization: An industrial control study.
\newblock {\em IEEE Transactions on Control Systems Technology}, 28:730--740, 2020.

\bibitem{Snoek2012PracticalBO}
Jasper Snoek, H.~Larochelle, and Ryan~P. Adams.
\newblock Practical bayesian optimization of machine learning algorithms.
\newblock In {\em NIPS}, 2012.

\bibitem{BOOO}
Fernando Nogueira.
\newblock {Bayesian Optimization}: Open source constrained global optimization tool for {Python}, 2014--.

\bibitem{Knig2021SafeAE}
Christopher K{\"o}nig, Matteo Turchetta, John Lygeros, Alisa Rupenyan, and Andreas Krause.
\newblock Safe and efficient model-free adaptive control via bayesian optimization.
\newblock {\em 2021 IEEE International Conference on Robotics and Automation (ICRA)}, pages 9782--9788, 2021.

\bibitem{Zhao2017PIDCD}
Cheng Zhao and Lei Guo.
\newblock Pid controller design for second order nonlinear uncertain systems.
\newblock {\em Science China Information Sciences}, 60:1--13, 2017.

\bibitem{Andrade2021UnmannedAV}
Fabio A.~A. Andrade, Ihannah~Pinto Guedes, Guilherme~F. Carvalho, Alessandro R.~L. Zachi, Diego~Barreto Haddad, Luciana~Faletti Almeida, Aur{\'e}lio~G. de~Melo, and Milena~Faria Pinto.
\newblock Unmanned aerial vehicles motion control with fuzzy tuning of cascaded-pid gains.
\newblock {\em Machines}, 2021.

\bibitem{Bhatia1970StabilityTO}
Nam~P. Bhatia and G.~P. Szeg{\"o}.
\newblock Stability theory of dynamical systems.
\newblock 1970.

\bibitem{Feler1995APO}
Robert Fe{\ss}ler.
\newblock A proof of the two-dimensional markus-yamabe stability conjecture and a generalization.
\newblock {\em Annales Polonici Mathematici}, 62:45--74, 1995.

\bibitem{KUZNETSOV2018138}
N.V. Kuznetsov, O.A. Kuznetsova, D.~Koznov, R.N. Mokaev, and B.~Andrievsky.
\newblock Counterexamples to the kalman conjectures.
\newblock {\em IFAC-PapersOnLine}, 51(33):138--143, 2018.
\newblock 5th IFAC Conference on Analysis and Control of Chaotic Systems CHAOS 2018.

\bibitem{669023}
K.~Tanaka, T.~Ikeda, and H.O. Wang.
\newblock Fuzzy regulators and fuzzy observers: relaxed stability conditions and lmi-based designs.
\newblock {\em IEEE Transactions on Fuzzy Systems}, 6(2):250--265, 1998.

\bibitem{481841}
H.O. Wang, K.~Tanaka, and M.F. Griffin.
\newblock An approach to fuzzy control of nonlinear systems: stability and design issues.
\newblock {\em IEEE Transactions on Fuzzy Systems}, 4(1):14--23, 1996.

\bibitem{4565671}
H.~K. Lam and Mohammad Narimani.
\newblock Stability analysis and performance design for fuzzy-model-based control system under imperfect premise matching.
\newblock {\em IEEE Transactions on Fuzzy Systems}, 17(4):949--961, 2009.

\bibitem{iet:/content/journals/10.1049/iet-cta.2010.0619}
H.K. Lam.
\newblock Stability analysis of polynomial fuzzy-model-based control systems under perfect/imperfect premise matching.
\newblock {\em IET Control Theory \& Applications}, 5:1689--1697(8), October 2011.

\bibitem{Lam2016PolynomialFM}
Hak~Keung Lam.
\newblock Polynomial fuzzy model-based control systems: Stability analysis and control synthesis using membership function dependent techniques.
\newblock 2016.

\bibitem{BANSAL2021375}
Nishtha Bansal, Aman Bisht, Sruti Paluri, Vineet Kumar, K.P.S. Rana, Ahmad~Taher Azar, and Sundarapandian Vaidyanathan.
\newblock Chapter 15 - single-link flexible joint manipulator control using backstepping technique.
\newblock In Sundarapandian Vaidyanathan and Ahmad~Taher Azar, editors, {\em Backstepping Control of Nonlinear Dynamical Systems}, Advances in Nonlinear Dynamics and Chaos (ANDC), pages 375--406. Academic Press, 2021.

\bibitem{gym}
Greg Brockman, Vicki Cheung, Ludwig Pettersson, Jonas Schneider, John Schulman, Jie Tang, and Wojciech Zaremba.
\newblock Openai gym, 2016.

\end{thebibliography}

\end{document}